\documentclass[english]{article}
\PassOptionsToPackage{natbib=true}{biblatex}
\usepackage[T1]{fontenc}
\usepackage{color}
\usepackage{babel}
\usepackage{array}
\RequirePackage{amsthm,amsmath,amsfonts,amssymb}
\usepackage{graphicx}
\usepackage{setspace}
\usepackage{caption} 
\usepackage[unicode=true,pdfusetitle,
 bookmarks=true,bookmarksnumbered=false,bookmarksopen=false,
 breaklinks=false,pdfborder={0 0 0},pdfborderstyle={},backref=false,colorlinks=true]
 {hyperref}

\makeatletter
\providecommand{\tabularnewline}{\\}

\theoremstyle{plain}

\newtheorem{theorem}{Theorem}[section]
\newtheorem{lemma}[theorem]{Lemma}

\theoremstyle{definition}
\newtheorem{definition}[theorem]{Definition}
\newtheorem{example}{Example}
\newtheorem*{scenarioCC1}{Scenario C.C.1}
\newtheorem*{scenarioCC2}{Scenario C.C.2}
\newtheorem*{scenarioCM1}{Scenario C.M.1}
\newtheorem*{scenarioCM2}{Scenario C.M.2}
\newtheorem*{scenarioBC1}{Scenario B.C.1}
\newtheorem*{scenarioBC2}{Scenario B.C.2}
\newtheorem*{scenarioBM1}{Scenario B.M.1}
\newtheorem*{scenarioBM2}{Scenario B.M.2}

\theoremstyle{remark}
\newtheorem*{remark}{Remark}

\newcommand{\beginsupplement}{%
        \setcounter{table}{0}
        \renewcommand{\thetable}{S.\arabic{table}}%
        \setcounter{figure}{0}
        \renewcommand{\thefigure}{S.\arabic{figure}}%
        \setcounter{section}{0}
        \renewcommand{\thesection}{S.\arabic{section}}%
        \setcounter{example}{0}
        \renewcommand{\theexample}{S.\arabic{example}}%
}

\@ifundefined{date}{}{\date{}}
\usepackage{fullpage}
\usepackage{algorithm, algorithmic}
\usepackage{colortbl}

\let\proglang=\textsf
\newcommand{\pkg}[1]{{\fontseries{m}\fontseries{b}\selectfont #1}}

\allowdisplaybreaks

\makeatother

\usepackage[style=authoryear]{biblatex}

\begin{document}

\title{Variable Selection Using Bayesian Additive Regression Trees}
\author{Chuji Luo\footnote{Chuji Luo is Data Scientist, Google LLC, Mountain View, California 94043,  USA (email: {\tt cjluo@ufl.edu}).} \and Michael J. Daniels\footnote{Michael J. Daniels is Professor and Chair, Department of Statistics, University  of Florida, Gainesville, Florida 32611, USA (email: {\tt daniels@ufl.edu}).}}
\maketitle

\begin{abstract}
Variable selection is an important statistical problem. This problem
becomes more challenging when the candidate predictors are of mixed
type (e.g. continuous and binary) and impact the response variable
in nonlinear and/or non-additive ways. In this paper, we review existing
variable selection approaches for the Bayesian additive regression
trees (BART) model, a nonparametric regression model, which is flexible
enough to capture the interactions between predictors and nonlinear
relationships with the response. An emphasis of this review is on
the capability of identifying relevant predictors. We also propose
two variable importance measures which can be used in a permutation-based
variable selection approach, and a backward variable selection procedure
for BART. We present simulations demonstrating that our approaches
exhibit improved performance in terms of the ability to recover all
the relevant predictors in a variety of data settings, compared to
existing BART-based variable selection methods.
\end{abstract}
\section{Introduction\label{sec:1}}
Variable selection, also known as feature selection in machine learning,
is the process of selecting a subset of relevant variables for use
in model construction. It has been, and continues to be a major focus
of research and practice because researchers and practitioners often
seek low-cost, interpretable and not overfitted models. For example,
in causal inference, model-based estimation of the effects of an exposure
on an outcome is generally sensitive to the choice of confounders
included in the model. If there are no unmeasured confounders, the
full model with all the confounders and many non-confounders generally
yields an unbiased estimation, but with a large standard error (\citet{wang2012bayesian}).
On the contrary, a model with fewer non-confounders not only produces
an unbiased estimation with a smaller standard error, but also saves
computation time.

Variable selection is often carried out in parametric settings, such
as the linear regression model (\citet{efroymson1960multiple,george1993variable,
tibshirani1996regression,zou2005regularization,carvalho2010horseshoe,
bhattacharya2015dirichlet}).
However, variable selection approaches based on a linear model often
fail when the underlying relationship between the predictors and the
response variable is nonlinear and/or non-additive, and it is generally
challenging to extend them to nonparametric models, such as tree-based
models which incorporate both main effects and interaction effects
of varying orders.

Tree-based models have been developed from both the frequentist and
Bayesian perspectives, including but not limited to random forests
(\citet{breiman2001random}), stochastic gradient boosting (\citet{friedman2002stochastic}),
reinforcement learning trees (\citet{zhu2015reinforcement}) and Bayesian
additive regression trees (\citet{chipman2010bart}), the last of
which is the focus of this paper. Variable selection for the tree-based
models in machine learning is often achieved through variable importance.
Taking random forests as an example, there are two types of variable
importance measures typically used: Gini importance (\citet{friedman2001greedy})
and permutation importance (\citet{breiman2001random}). Gini importance
evaluates the importance of a variable by adding up the weighted decrease
in impurity of all the nodes using the variable as a split variable,
averaged over all the trees, while permutation importance evaluates
the importance of a variable by adding up the difference in out-of-bag
error before and after the permutation of the values of the variable
in the training data, also averaged over all the trees. Though widely
used, as pointed out by \citet{strobl2007bias}, both Gini importance
and permutation importance are biased in favor of continuous and high
cardinality predictors when the random forest is applied to data with
mixed-type predictors. Methods such as growing unbiased trees (\citet{strobl2007bias}) 
and partial permutations (\citet{altmann2010permutation}) 
can be used to solve this issue.

Compared to the tree-based models in machine learning, Bayesian additive
regression trees (BART) not only possesses competitive predictive
performance, but also enables the user to make inference on it due
to its fully Bayesian construction. The original BART paper suggests
using variable inclusion proportions to measure the importance of
predictors and later \citet{bleich2014variable} propose a principled
permutation-based inferential approach to determine how large the
variable inclusion proportion has to be in order to select a predictor.
This approach exhibits superior performance in a variety of data settings,
compared to many existing variable selection procedures including
random forests with permutation importance. In fact, we note that the
variable inclusion proportion produced by BART is a special case of
Gini importance, which treats the weight and the decrease in impurity
at each node to be constant. We also note that the approach of \citet{bleich2014variable}
is reminiscent of the partial permutation approach of \citet{altmann2010permutation}
which also repeatedly permutes the response vector to estimate the
distribution of the variable importance for each predictor in a non-informative
setting where the relationship between the response and predictors
is removed and the dependencies among the predictors are maintained.
Though the approach of \citet{altmann2010permutation} is designed
for mitigating the bias of Gini importance, we find that the approach
of \citet{bleich2014variable} is still biased against predictors
with low cardinality, such as binary predictors, especially when the
number of predictors is not large. In addition to the approach of
\citet{bleich2014variable}, two variable selection approaches are
proposed based on variants of the BART model. One variant of BART
is DART (\citet{linero2018bayesian}) which modifies BART by placing
a Dirichlet hyper-prior on the splitting proportions of the regression
tree prior to encourage sparsity. To conduct variable selection, they
suggest selecting predictors in the median probability model (\citet{barbieri2004optimal}),
i.e., selecting predictors with a marginal posterior variable inclusion
probability at least $50\%$. This approach is more computationally
efficient than other BART-based variable selection methods, as it
does not require fitting model multiple times. In general, we find
that DART works better than other BART-based variable selection methods,
but it becomes less stable in presence of correlated predictors and
in the probit BART model. Another variant of the BART prior is the
spike-and-forest prior (\citet{rovckova2020posterior}) which wraps
the BART prior with a spike-and-slab prior on the model space. \citet{liu2019variable}
provide model selection consistency results for the spike-and-forest
prior and propose approximate Bayesian computation (ABC) Bayesian
forest, a modified ABC sampling method, to conduct variable selection
in practice. Similar to DART, variables are selected based on their
marginal posterior variable inclusion probabilities. ABC Bayesian
forest shows good performance in terms of excluding irrelevant predictors,
but its ability to include relevant predictors is relatively poor
when predictors are correlated.

The main issues of existing BART-based variable selection methods
include being biased against categorical predictors with fewer levels
and being conservative in terms of including relevant predictors in
the model. The goal of this paper is to develop variable selection
approaches for BART to overcome these issues. For simplicity, we assume
that the possible types of potential predictors only include binary
and continuous types. This paper is organized as follows. In Section
\ref{sec:2}, we review the BART model and existing BART-based variable
selection approaches. An example where the approach of \citet{bleich2014variable}
fails to include relevant binary predictors is also provided. In Section
\ref{sec:3}, we present the proposed variable selection approaches.
In Section \ref{sec:4}, we compare our approaches with existing
BART-based variable selection approaches through simulated examples.
Finally, we conclude this paper with a discussion in Section \ref{sec:5}.

\section{Review of BART\label{sec:2}}

\subsection{Model Specification\label{subsec:2_1}}

Motivated by the boosting algorithm and built on the Bayesian classification
and regression tree algorithm (\citet{chipman1998bayesian}), BART (\citet{chipman2010bart})
is a Bayesian approach to nonparametric function estimation and inference
using a sum of trees. Consider the problem of making inference about
an unknown function $f_{0}$ that predicts a continuous response $y$
using a $p$-dimensional vector of predictors 
$\boldsymbol{x}=(x_{1},\cdots,x_{p})^{\intercal}$ when 
\begin{equation}
y = f_{0}(\boldsymbol{x}) + \epsilon,\quad \epsilon\sim N(0,\sigma^{2}).\label{eq:2_1}
\end{equation}
BART models $f_{0}$ by a sum of $M$ Bayesian regression trees, i.e.,
\begin{equation}
f\left(\boldsymbol{x}\right) = \sum\limits _{m=1}^{M} g\left(\boldsymbol{x}; T_{m}, 
\boldsymbol{\mu}_{m} \right),\label{eq:2_2}
\end{equation}
where $g(\boldsymbol{x};T_{m},\boldsymbol{\mu}_{m})$ is the output
of $\boldsymbol{x}$ from a single regression tree. Each 
$g(\boldsymbol{x};T_{m},\boldsymbol{\mu}_{m})$
is determined by the binary tree structure $T_{m}$ consisting of
a set of splitting rules and a set of terminal nodes, and the vector
of parameters $\boldsymbol{\mu}_{m}=(\mu_{m,1},\cdots,\mu_{m,b_{m}})$
associated with the $b_{m}$ terminal nodes of $T_{m}$, such that
$g(\boldsymbol{x};T_{m},\boldsymbol{\mu}_{m})=\mu_{m,l}$ if $\boldsymbol{x}$
is associated with the $l^{th}$ terminal node of the tree $T_{m}$.

A regularization prior consisting of three components, (1) the $M$
independent trees structures $\{T_{m}\}_{m=1}^{M}$, (2) the parameters
$\{\boldsymbol{\mu}_{m}\}_{m=1}^{M}$ associated with the terminal
nodes given the trees $\{T_{m}\}_{m=1}^{M}$ and (3) the error variance
$\sigma^{2}$ which is assumed to be independent of the former two,
is specified for BART in a hierarchical manner. The posterior distribution
of BART is sampled through a Metropolis-within-Gibbs algorithm (\citet{hastings1970monte,geman1984stochastic,kapelner2013bartmachine}).
The Gibbs sampler involves $M$ successive draws of $(T_{m},\boldsymbol{\mu}_{m})$
followed by a draw of $\sigma^{2}$. A key feature of this sampler
is that it employs Bayesian backfitting (\citet{hastie2000bayesian})
to sample each pair of $(T_{m},\boldsymbol{\mu}_{m})$.

BART estimates $f_{0}(\boldsymbol{x})=E(Y\mid\boldsymbol{x})$ by
taking the average of the posterior samples of $f(\boldsymbol{x})$
after a burn-in period. As the trees structures are being updated
through the MCMC chain, the model space is being searched by BART.
In light of this, the estimation of $f_{0}(\boldsymbol{x})$ can be
considered as a model averaging estimation with each posterior sample
of the trees structures treated as a model and therefore BART can
be regarded as a Bayesian model selection approach.

The continuous BART model (\ref{eq:2_1}--\ref{eq:2_2}) can be extend
to a binary regression model for a binary response $Y$ by specifying
\begin{equation}
P\left(Y=1\mid\boldsymbol{x}\right) = \Phi\left[f\left(\boldsymbol{x}\right)\right],\label{eq:2_3}
\end{equation}
where $f$ is the sum-of-trees function in (\ref{eq:2_2}) and $\Phi$
is the cumulative distribution function of the standard normal distribution.
Posterior sampling can be done by applying the Metropolis-within-Gibbs
sampler of continuous BART to the latent variables which are obtained
via the data augmentation approach of \citet{albert1993bayesian}.

\subsection{Existing BART-Based Variable Selection Methods\label{subsec:2_2}}

In this section, we review the BART variable inclusion proportion,
a variable importance measure produced by BART, and three existing
variable selection methods based on BART.

\subsubsection{BART variable inclusion proportions\label{subsec:2_2_1}}

The MCMC algorithm of BART returns posterior samples of the trees
structures, from which we can obtain the variable inclusion proportion
for each predictor.

\begin{definition}
[BART Variable Inclusion Proportion\label{def:2_1}] Let $K$ be the
number of posterior samples obtained from a BART model. For each $j=1,\cdots,p$
and $k=1,\cdots,K$, let $c_{jk}$ be the number of splitting rules
using the predictor $x_{j}$ as the split variable in the $k^{\text{th}}$
posterior sample of the trees structures and let $c_{\cdot k}=\sum_{j=1}^{p}c_{jk}$
be the total number of splitting rules in the $k^{\text{th}}$ posterior
sample. The BART variable inclusion proportion
(VIP) for the predictor $x_{j}$ is defined as
\begin{equation*}
v_{j} = \frac{1}{K}\sum\limits _{k=1}^{K}\frac{c_{jk}}{c_{\cdot k}}.
\end{equation*}
\end{definition}

The BART VIP $v_{j}$ describes the average usage per splitting rule
for the predictor $x_{j}$. Intuitively, a large VIP is suggestive
of the predictor being an important driver of the response.
Therefore, the original BART paper uses
VIPs to rank predictors in terms of relative importance and recommends
conducting variable selection using VIPs with a small number of trees
in order to introduce competition among predictors.

We note that BART VIP is a special case of Gini importance (\citet{louppe2014understanding})
which evaluates the importance of a variable by adding up the weighted
decrease in impurity of all the nodes using the variable as a split
variable, averaged over all the trees in the ensemble:
\begin{equation*}
\text{IMP}_{\text{Gini}}\left(x_{j}\right) = \frac{1}{M}\sum_{m=1}^{M}
\sum_{\eta\in\phi_{m}}\mathbb{I}\left(j_{\eta}=j\right)\left[w(\eta)\Delta i(\eta)\right],
\end{equation*}
where $\phi_{m}$ is the set of all the splitting nodes in the $m^{\text{th}}$
tree, $j_{\eta}$ is the index of the variable used for the splitting
node $\eta$, $w(\eta)$ is the weight of the node $\eta$ and $\Delta i(\eta)$
is the decrease in impurity of the node $\eta$. In random forests,
the weight $w(\eta)$ is typically the proportion of the samples reaching
the node $\eta$ and the impurity $i(\eta)$ is typically the variance
of the observed response values of the samples reaching the node $\eta$.
BART VIP can be regarded as Gini importance with the weight $w(\eta)$
set to $1/c_{\cdot k}$ and the impurity $i(\eta)$ set to $1$ for
all the nodes.

Gini importance is known to be biased against variables with low cardinality
in random forests (\citet{strobl2007bias}). As a variant of Gini
importance, BART VIP is also a biased variable importance measure.
To illustrate this, we consider the following example.

\begin{example}\label{exp:1}
For each $i=1, \cdots, 500$, sample $x_{i1}, \cdots, x_{i10}$ from $\text{Bernoulli(0.5)}$
independently, $x_{i11}, \cdots, x_{i20}$ from $\text{Uniform(0,1)}$ independently and $y_{i}$
from $\mathrm{Normal}(f_{0}(\boldsymbol{x}_{i}),1)$ independently, where
\begin{equation*}
f_{0}(\boldsymbol{x}) = 10\sin(\pi x_{1}x_{11})+20(x_{13}-0.5)^{2}+10x_{2}+5x_{12}.
\end{equation*}
\end{example}

\begin{figure}[ht]
\centering\includegraphics[width=0.7\columnwidth]{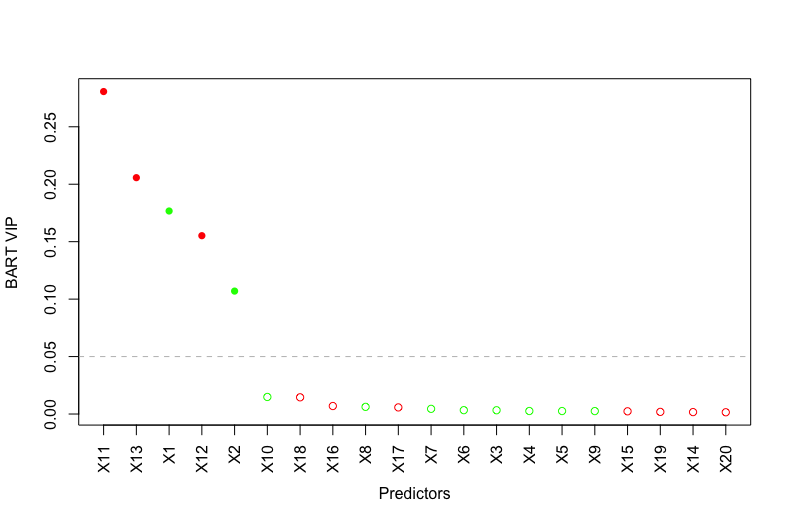}
\caption{\label{fig:BART-VIP}BART VIPs for different predictors in
Example \ref{exp:1}. Red dots are for continuous predictors
and green dots are for binary predictors.}
\end{figure}

This example is a variant of Friedman's example (\citet{friedman1991multivariate})
where the predictors impact the response in a nonlinear and non-additive
way. We modify Friedman's example by changing the predictors $x_{1}$
and $x_{2}$ from continuous to binary. We fit a BART model on the
observations $\{y_{i},\boldsymbol{x}_{i}\}_{i=1}^{500}$ with $M=20$
trees and obtain the VIPs based on $1000$ posterior samples after
burning in $1000$ posterior samples. Since all the predictors are
on the same scale $[0,1]$, the
binary predictor $x_{1}$ improves the fit of the response $y$ the
same as the continuous predictor $x_{11}$ and the binary predictor
$x_{2}$ moves the response $y$ more than the continuous predictor
$x_{12}$. However, Figure \ref{fig:BART-VIP} shows that $x_{11}$
and $x_{12}$ have higher VIPs than $x_{1}$ and $x_{2}$ respectively,
suggesting that BART VIP is biased against binary predictors.
This issue occurs because a binary
predictor has only one available splitting value while a continuous
predictor has many more splitting values. As a result, continuous
predictors appear more often in the ensemble.

\subsubsection{Permutation-based variable selection with BART VIP\label{subsec:2_2_2}}

Apart from the bias issue of BART VIP, the original BART paper does
not provide a complete variable selection approach in the sense that
it does not provide a way of setting the threshold for BART VIPs to
select predictors. In fact, \citet{bleich2014variable} show that
it is not possible to decide on an appropriate threshold for VIPs
based on a single BART fit, because BART may fit noise due to its
nonparametric flexibility. To deal with this, they develop a principled
permutation-based variable selection approach which facilitates determining
how large a BART VIP has to be in order to select a predictor.

Specifically, this approach first creates $L$ permutations 
$\{\boldsymbol{y}_{l}^{*}\}_{l=1}^{L}$ of the response vector $\boldsymbol{y}$ 
to form $L$ null datasets $\{\boldsymbol{y}_{l}^{*},X\}_{l=1}^{L}$, 
where $X$ is a matrix of the observed predictors with each row 
corresponding to an observation, and then fits a BART model on each null dataset 
$(\boldsymbol{y}_{l}^{*},X)$. As a result, $L$ vectors of BART VIPs 
$\boldsymbol{v}_{l}^{*}=(v_{l,1}^{*},\cdots,v_{l,p}^{*})$,
$l=1,\cdots,L$, can be obtained from the $L$ BART models. Since
the null datasets delete the relationship between the response and
predictors, the $L$-dimensional vector $\{v_{1,j}^{*},\cdots,v_{L,j}^{*}\}$
can be regarded as a sample of the null distribution of the BART VIP
for the predictor $x_{j}$, which is the distribution of the BART
VIP when the predictor $x_{j}$ is unrelated to the response $y$.
If the predictor $x_{j}$ is indeed not related to the response $y$,
then the true BART VIP obtained from the BART model on the original
dataset $(\boldsymbol{y},X)$ follows the null distribution. Hence,
predictors can be selected according to the relative location of their
true BART VIPs in the corresponding null distributions. Since BART
may fit noise, the true VIPs are estimated by the averaged VIPs 
$\bar{\boldsymbol{v}}=(\bar{v}_{1},\cdots,\bar{v}_{p})$
which is the mean of multiple vectors of VIPs with each obtained from
a replication of BART on the original dataset $(\boldsymbol{y},X)$.
\citet{bleich2014variable} propose three criteria to select a subset
of predictors. The least stringent one is the local threshold, i.e.,
the predictor $x_{j}$ is selected if the averaged VIP $\bar{v}_{j}$
exceeds the $1-\alpha$ quantile of the corresponding empirical distribution
$\sum_{l=1}^{L}\delta_{v_{l,j}^{*}}(\cdot)$ of the null distribution,
where $\delta_{v_{l,j}^{*}}(\cdot)$ is a degenerate distribution
at $v_{l,j}^{*}$. The algorithm for a general permutation-based variable
selection approach, not restricted to the one using BART VIP as the
variable importance measure, is summarized in Algorithm \ref{alg:2_1}.

\begin{algorithm}[ht]
\begin{algorithmic}[1]
\REQUIRE {$X$: predictors; $\boldsymbol{y}$: response; $L$: number of  permutations; $L_{\text{rep}}$: number of repetitions; $\alpha$: selection threshold}
\ENSURE{ A subset of predictors}
\FOR {$s=1,\cdots,L_{\text{rep}}$}
\STATE Run the model on the original dataset $(\boldsymbol{y},X)$ and compute the variable importance scores $(v_{s,1}, \cdots, v_{s,p})$
\ENDFOR
\STATE Compute the averaged  variable importance scores $\bar{v}_j =  (\sum_{s=1}^{L_{\text{rep}}} v_{s,j})/L_{\text{rep}}$ for $j=1,\cdots,p$
\FOR {$l=1,\cdots,L$}
\STATE Run the model on the null dataset $(\boldsymbol{y}^*_l,X)$ and compute the variable importance scores $( v_{l,1}^{*},\cdots,v_{l,p}^{*} )$
\ENDFOR
\FOR {$j=1,\cdots,p$}
\STATE Compute the $1-\alpha$ quantile $v_j^{\alpha}$ of the  empirical distribution  $\sum_{l=1}^L \delta_{v^*_{l,j}}(\cdot)$ for the predictor $x_j$
\IF {$\bar{v}_j > v_j^{\alpha}$}
\STATE Select $x_j$
\ENDIF
\ENDFOR
\end{algorithmic} 
\caption{Permutation-based variable selection approach}\label{alg:2_1}
\end{algorithm}

This approach works well when all the predictors are continuous; however,
it often fails to include relevant binary predictors when the predictors
belong to different types and the number of predictors is small. Consider
the following variant of Friedman's example.

\begin{example}\label{exp:2}
For each $i=1, \cdots, 500$, sample $x_{i1}, \cdots, x_{i10}$ from $\text{Bernoulli(0.5)}$
independently, $x_{i11}, \cdots, x_{i20}$ from $\text{Uniform(0,1)}$ independently and $y_{i}$
from $\mathrm{Normal}(f_{0}(\boldsymbol{x}_{i}),1)$ independently, where
\begin{equation*}
f_{0}(\boldsymbol{x}) = 10\sin(\pi x_{11}x_{12})+20(x_{13}-0.5)^{2}+10x_{1}+5x_{2}.
\end{equation*}
\end{example}

Figure S.1 of the Supplementary Material shows that both the two relevant 
binary predictors $x_{1}$ and $x_{2}$ are not selected by this method. 
Further analysis is provided in Section \ref{subsec:3_1}.

\subsubsection{Variable selection with DART\label{subsec:2_2_3}}

DART (\citet{linero2018bayesian}) is a variant of BART, which replaces
the discrete uniform distribution for selecting a split variable with
a categorical distribution of which the event probabilities $(s_{1},\cdots,s_{p})$
follow a Dirichlet distribution. With the Dirichlet hyper-prior, probability
mass is accumulated towards the predictors that are more frequently
used along the MCMC iterations, thereby inducing sparsity and improving
the predictive performance in high dimensional settings.

As discussed in Section \ref{subsec:2_1}, BART can be considered
as a Bayesian model selection approach, so predictors can be selected
by using the marginal posterior variable inclusion probability (MPVIP)
\begin{equation}
\pi_{j}=P(x_{j}\text{ in the model}\mid\boldsymbol{y})\label{eq:2_4}. 
\end{equation}
However, the model space explored by BART is not only all the possible combinations
of the predictors, but also all the possible relationships that the
sum-of-trees model can express using the $p$ predictors, which implies
that the cardinality of the model space is much bigger than $2^{p}$
and that the variable selection results can be susceptible to the
influence of noise predictors, especially when the number of predictors
is large. DART avoids the influence of noise predictors by employing
the shrinkage Dirichlet hyper-prior. DART estimates the MPVIP for
a predictor by the proportion of the posterior samples of the trees
structures where the predictor is used as a split variable at least
once, and selects predictors with MPVIP at least $0.5$, yielding
a median probability model. \citet{barbieri2004optimal} shows that
the median probability model, rather than the most probable model,
is the optimal choice in the sense that the predictions for future
observations are closest to the Bayesian model averaging prediction
in the squared error sense. However, the assumptions for this property
are quite strong and often not realistic, including the assumption
of an orthogonal predictors matrix.

\subsubsection{Variable selection with ABC Bayesian forest\label{subsec:2_2_4}}

\citet{rovckova2020posterior} introduce a spike-and-forest prior
which wraps the BART prior with a spike-and-slab prior on the model
space:
\begin{align*}
& \,\, \mathcal{S} \sim \pi\left(\mathcal{S}\right), \, \forall \mathcal{S}\subseteq\left\{ 1,\cdots,p\right\}, \\
& \left\{ T_{m},\boldsymbol{\mu}_{m}\right\} _{m=1}^{M},\sigma^{2}\mid\mathcal{S} \sim \text{BART prior}.
\end{align*}
The tree-based model under the spike-and-forest prior can be used not only 
for estimation but also for variable selection,
as shown in \citet{liu2019variable}. Due to the intractable marginal
likelihood, they propose an approximate Bayesian computation (ABC)
sampling method based on data-splitting to help sample from the model
space with higher ABC acceptance rate. Specifically, at each iteration,
the dataset is randomly split into a training set and a test set according
to a certain split ratio. The algorithm proceeds by sampling a subset
$\mathcal{S}$ from the prior $\pi(\mathcal{S})$, fitting a BART
model on the training set only with the predictors in $\mathcal{S}$,
and computing the root mean squared errors (RMSE) for the test set
based on a posterior sample from the fitted BART model. Only those
subsets that result in a low RMSE on the test set are kept for selection.
Similar to DART, ABC Bayesian forest selects predictors based on their
MPVIP $\pi_{j}$ defined as (\ref{eq:2_4}) which is estimated by computing 
the proportion of ABC accepted BART posterior samples that use 
the predictor $x_{j}$ at least one time. Given the $\pi_{j}$'s, 
predictors with $\pi_{j}$ exceeding a pre-specified threshold are selected. 

ABC Bayesian forest exhibits good performance in excluding irrelevant
predictors, but it does not perform well in including relevant predictors
in presence of correlated predictors, as shown in Section \ref{sec:4}. 
One possible reason is that ABC Bayesian forest internally uses only one 
posterior sample after a short burn-in period and a small number of trees 
for BART, making it easy to miss relevant predictors. 
Furthermore, given that ABC Bayesian forest computes $\pi_{j}$ based on 
the ABC accepted BART posterior samples rather than the ABC accepted subsets, 
it appears that the good performance in excluding irrelevant predictors may be primarily
due to the variable selection capability of BART itself, when the
number of trees is small. 

\section{New Approaches\label{sec:3}}

\subsection{BART VIP with Type Information\label{subsec:3_1}}

In Section \ref{subsec:2_2_2}, we show an example
where the approach of \citet{bleich2014variable} fails to include
relevant binary predictors. The intuition of this approach is
that a relevant predictor is expected to appear more often in the
model built on the original dataset than that built on a null dataset.
In this section, we revisit Example \ref{exp:2} and show
that the intuition does hold for both relevant continuous and relevant
binary predictors, but that the increase in usage of a relevant binary
predictor is often offset by the increase of relevant continuous predictors,
when the number of predictors is not large, thereby making the true
VIP of a relevant binary predictor insignificant with respect to (w.r.t.) the corresponding
null VIPs. In light of this observation, we modify BART VIP with the
type information of predictors to help identify relevant binary predictors.

Direct studying BART VIP is challenging because each $v_j$ consists of $K$
different components $\{c_{jk} / c_{\cdot k} \}_{k=1}^K$. 
To make the analysis more interpretable, we introduce an approximation of
BART VIP, which only contains one component.

\begin{definition}
[Approximation of BART VIP\label{def:3_1}]Following the notation
in Definition \ref{def:2_1}, for every $j=1,\cdots,p$ and $k=1,\cdots,K$,
let $c_{j\cdot}=\sum_{k=1}^{K}c_{jk}$ be the number of splitting
nodes using the predictor $x_{j}$ over all the posterior samples
of the trees structures and $c_{\cdot\cdot}=\sum_{j,k=1}^{p,K}c_{jk}$ 
be the total number of splitting nodes over all the posterior samples. 
For each predictor $x_j$, define
\begin{equation*}
\tilde{v}_{j} = \frac{c_{j\cdot}}{c_{\cdot\cdot}}.
\end{equation*}
\end{definition}

The following lemma shows that $\tilde{v}_{j}$ defined above
can be used to approximate BART VIP $v_j$ under mild conditions.

\begin{lemma}\label{lem:3_1}
For any $j=1,\cdots,p$ and $k=1,\cdots,K$,
let $c_{\cdot k}=\sum_{j=1}^{p}c_{jk}$ be the
total number of splitting nodes in the $k^{\text{th}}$ posterior
sample and $\bar{c}_{\cdot K}=c_{\cdot\cdot}/K$
be the average number of splitting nodes in a posterior sample.
If $\sum_{k=1}^{K}(c_{jk}/c_{\cdot k})^{2}/K \le \delta_{1}$ and 
$[\sum_{k=1}^{K}(c_{\cdot k}-\bar{c}_{\cdot K})^{2}/K]^{1/2}/\bar{c}_{\cdot K}\le\delta_{2}$
for some positive numbers $\delta_{1},\delta_{2}>0$, then the difference
between $\tilde{v}_{j}$ and the corresponding BART VIP $v_{j}$ is
bounded, i.e., $\lvert\tilde{v}_{j}-v_{j}\rvert\le\delta_{1}^{1/2}\delta_{2}$.
\end{lemma}

\begin{remark}
The first condition above means that the variance of $c_{jk}/c_{\cdot k}$, 
the proportion of the usage of the predictor
$x_{j}$ in an ensemble, is bounded. The second condition means that the coefficient
of variation of $c_{\cdot k}$, the number of splitting nodes in an ensemble, 
is also bounded. Since \citet{bleich2014variable} use posterior samples 
after a burn-in period, the variance of $c_{jk}/c_{\cdot k}$
and the coefficient of variation of $c_{\cdot k}$ are small (see
Figure S.2 of the Supplementary Material). 
Hence, Lemma \ref{lem:3_1} implies that $\tilde{v}_{j}$ can be used as 
an alternative to the corresponding BART VIP $v_{j}$.
\end{remark}

\begin{proof}
See Section S.1 of the Supplementary Material.
\end{proof}

Consider the approach of \citet{bleich2014variable}. 
For every $j=1,\cdots,p$, $k=1,\cdots,K$, $r=1,\cdots,L_{\text{rep}}$ 
and $l=1,\cdots,L$, let $c_{l,jk}^{*}$ be the number of splitting nodes 
using the predictor $x_{j}$ in the $k^{\text{th}}$ posterior sample of 
the BART model built on the $l^{\text{th}}$ null dataset and 
$c_{r,jk}$ be that of the $r^{\text{th}}$ repeated BART model built on 
the original dataset. 
Following the notation in Definition \ref{def:3_1}, 
write $c_{l,j\cdot}^{*}=\sum_{k=1}^{K}c_{l,jk}^{*}$,
$c_{l,\cdot\cdot}^{*}=\sum_{j,k=1}^{p,K}c_{l,jk}^{*}$, 
$c_{r,j\cdot}=\sum_{k=1}^{K}c_{r,jk}$,
and $c_{r,\cdot\cdot}=\sum_{j,k=1}^{p,K}c_{r,jk}$. 
By Lemma \ref{lem:3_1}, for each predictor $x_{j}$, 
we can use $\tilde{v}_{lj}^{*}=c_{l,j\cdot}^{*}/c_{l,\cdot\cdot}^{*}$
to approximate $v_{lj}^{*}$, the VIP obtained from the BART model
on the $l^{\text{th}}$ null dataset, and $\tilde{v}_{rj}=c_{r,j\cdot}/c_{r,\cdot\cdot}$
to approximate $v_{rj}$, the VIP obtained from the $r^{\text{th}}$
repeated BART model on the original dataset. 
Denote by $\bar{\tilde{v}}_{j}=(\sum_{r=1}^{L_{\text{rep}}}\tilde{v}_{rj})/L_{\text{rep}}$
the averaged approximate VIP for the predictor $x_{j}$ across $L_{\text{rep}}$
repetitions of BART on the original dataset. 
According to the local threshold of \citet{bleich2014variable}, 
a predictor $x_{j}$ is selected only if the average VIP 
$\bar{\tilde{v}}_{j}$ exceeds the $1-\alpha$ quantile of the empirical distribution 
$\sum_{l=1}^{L}\delta_{\tilde{v}_{lj}^{*}}(\cdot)$, i.e., 
\begin{equation}
\frac{1}{L}\sum\limits _{l=1}^{L}\mathbb{I}\left(\frac{1}{L_{\text{rep}}}\sum\limits _{r=1}^{L_{\text{rep}}}\frac{c_{r,j\cdot}}{c_{r,\cdot\cdot}}>\frac{c_{l,j\cdot}^{*}}{c_{l,\cdot\cdot}^{*}}\right)\,>\,1-\alpha.\label{eq:2_5}
\end{equation}
Similar to the approximation of BART VIP, we can approximate
$(\sum_{r=1}^{L_{\text{rep}}}c_{r,j\cdot}/c_{r,\cdot\cdot})/L_{\text{rep}}$
in (\ref{eq:2_5}) by $\bar{c}_{L_{\text{rep}},j\cdot}/\bar{c}_{L_{\text{rep}},\cdot\cdot}$
and therefore the selection criteria (\ref{eq:2_5}) can be rewritten as
\begin{equation}
\frac{1}{L}\sum\limits _{l=1}^{L}\mathbb{I}\left(\frac{\bar{c}_{L_{\text{rep}},j\cdot}}{c_{l,j\cdot}^{*}}>\frac{\bar{c}_{L_{\text{rep}},\cdot\cdot}}{c_{l,\cdot\cdot}^{*}}\right)\,>\,1-\alpha,\label{eq:2_6}
\end{equation}
where $\bar{c}_{L_{\text{rep}},j\cdot}=(\sum_{r=1}^{L_{\text{rep}}}c_{r,j\cdot})/L_{\text{rep}}$
and $\bar{c}_{L_{\text{rep}},\cdot\cdot}=(\sum_{r=1}^{L_{\text{rep}}}c_{r,\cdot\cdot})/L_{\text{rep}}$.

The ratio $\bar{c}_{L_{\text{rep}},j\cdot} / c_{l,j\cdot}^{*}$
on the left hand side (LHS) of the inequality inside the sum of (\ref{eq:2_6})
represents the average increment in usage of a predictor $x_{j}$
in a BART model built on the original dataset, compared to a null
dataset; the ratio $\bar{c}_{L_{\text{rep}},\cdot\cdot} / c_{l,\cdot\cdot}^{*}$
on the right hand side (RHS) represents the average increment in the
total number of the splitting nodes over all the posterior samples
of a BART model built on the original dataset, compared to a null
dataset. 
Figure S.3 of the Supplementary Material
shows the overall counts ratio
$\bar{c}_{L_{\text{rep}},\cdot\cdot} / c_{l,\cdot\cdot}^{*}$
and the counts ratio $\bar{c}_{L_{\text{rep}},j\cdot} / c_{l,j\cdot}^{*}$ 
for each predictor in Example \ref{exp:2}. 
Both relevant continuous ($x_{11}$, $x_{12}$
and $x_{13}$) and relevant binary ($x_{1}$ and $x_{2}$) predictors
are indeed used more frequently in the BART model built on the original
dataset, i.e., $\bar{c}_{L_{\text{rep}},j\cdot} / c_{l,j\cdot}^{*} > 1$,
but the increment in usage of a relevant binary predictor is
not always greater than the increment in the total number of splits,
i.e., $\bar{c}_{L_{\text{rep}},j\cdot} / c_{l,j\cdot}^{*} \ngtr \bar{c}_{L_{\text{rep}},\cdot\cdot} / c_{l,\cdot\cdot}^{*}$.
A binary predictor can only be used once at most in each path of a
tree because it only has one split value,
and a few splits are sufficient to provide the information about a
binary predictor. As a result, $\bar{c}_{L_{\text{rep}},j\cdot}$, the number of splits
using a binary predictor, is limited by its low cardinality, even if
it is relevant. In addition, the number of splits $c_{l,j\cdot}^{*}$
using a binary predictor in a BART model built on a null dataset is
roughly $1/p$ of the total number of splits in the ensemble
(\citet{bleich2014variable}). As such, when the number of predictors
is small, $c_{l,j\cdot}^{*}$ is close to $\bar{c}_{L_{\text{rep}},j\cdot}$
(see the left subfigure of Figure S.4 of the Supplementary 
Material), thereby making
the counts ratio $\bar{c}_{L_{\text{rep}},j\cdot} / c_{l,j\cdot}^{*}$
for a binary predictor close to $1$. The overall counts ratio $\bar{c}_{L_{\text{rep}},\cdot\cdot} / c_{l,\cdot\cdot}^{*}$
is also close to $1$ because of the BART regularization prior. In
fact, given a fixed number of trees in an ensemble, the total number
of splits in the ensemble is controlled by the splitting probability,
the probability of splitting a node into two child nodes, which is
the same for BART models built on the original dataset and null datasets,
so the total number of splits for the original dataset and a null
dataset, $\bar{c}_{L_{\text{rep}},\cdot\cdot}$ and $c_{l,\cdot\cdot}^{*}$,
are of similar magnitude (see the right subfigure of Figure S.4).
As such, the increment in usage of a relevant binary predictor 
$\bar{c}_{L_{\text{rep}},j\cdot} / c_{l,j\cdot}^{*}$ is easily offset by the increment in
the total number of splits $\bar{c}_{L_{\text{rep}},\cdot\cdot} / c_{l,\cdot\cdot}^{*}$.

Figure S.4 of the Supplementary Material also shows that 
the increment in the total number of splits is primarily driven by those 
relevant continuous predictors, so essentially the increment in usage of 
a relevant binary predictor is offset by that of relevant continuous predictors. 
To alleviate this behavior, we modify Definition \ref{def:2_1} using the type
information of predictors.

\begin{definition}
[BART Within-Type Variable Inclusion Proportion\label{def:3_2}]
Denote by $\mathcal{S}_{\text{cts}}$ (or $\mathcal{S}_{\text{bin}}$) the
set of indicators for continuous (or binary) predictors. Following
the notation in Definition \ref{def:2_1}, for every $k=1,\cdots,K$,
let $c_{\text{cts,}k}=\sum_{j\in\mathcal{S}_{\text{cts}}}c_{jk}$
(or $c_{\text{bin,}k}=\sum_{j\in\mathcal{S}_{\text{bin}}}c_{jk}$)
be the total number of splitting nodes using continuous (or binary)
predictors in the $k^{\text{th}}$ posterior sample. For every $j=1,\cdots,p$,
define the BART within-type VIP for the predictor $x_{j}$ as follows:
\begin{equation*}
v_{j}^{\text{w.t.}}\,=\,\frac{1}{K}\sum\limits _{k=1}^{K}\frac{c_{jk}}{c_{\text{cts,}k}\cdot\mathbb{I}(j\in\mathcal{S}_{\text{cts}})+c_{\text{bin,}k}^{\text{}}\cdot\mathbb{I}(j\in\mathcal{S}_{\text{bin}})}.
\end{equation*}
\end{definition}

\begin{figure}[ht]
\centering\includegraphics[width=0.7\columnwidth]{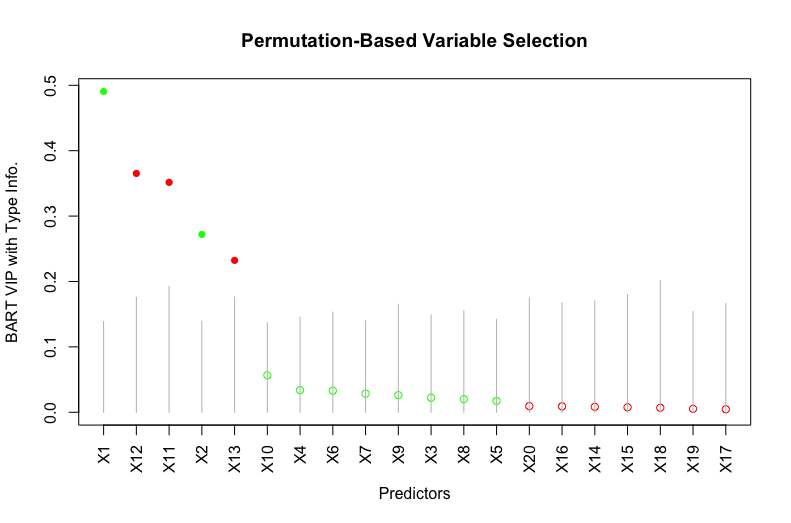}

\caption{\label{fig:within-type-Bleich}Variable selection results of the
permutation-based approach using BART within-type VIP as the variable
importance measure for Example \ref{exp:2}. The same setting 
($L=100$, $L_{\text{rep}}=10$, $\alpha=0.05$ and $M=20$)
and BART posterior samples as Figure S.1 of the Supplementary 
Material are used. Red (or green) dots are for continuous 
(or binary) predictors. Solid (or open) dots are for selected (or not selected) 
predictors. Each vertical grey line is the local threshold for the corresponding predictor.
Both the two relevant binary predictors, $x_{1}$ and $x_{2}$, 
are selected and have similar variable importance to
the relevant continuous ones.}
\end{figure}

Define $\delta_{j j^{\prime}} = \mathbb{I}(x_{j}\text{ and }x_{j^{\prime}}\text{ are of the same type})$
for any $j = 1, \cdots, p$ and $j^{\prime} = 1, \cdots, p$.
Write $\bar{c}_{L_{\text{rep}},\text{type of }x_{j},\cdot} = 
(\sum_{r=1}^{L_{\text{rep}}}\sum_{j^{\prime}=1}^{p}
c_{r,j^{\prime}\cdot} \cdot \delta_{j j^{\prime}}) / L_{\text{rep}}$ and 
$c_{l,\text{type of }x_{j},\cdot}^{*} = \sum_{j^{\prime}=1}^{p}
c_{l,j^{\prime}\cdot}^{*}  \cdot \delta_{j j^{\prime}}$. 
The permutation-based variable selection
approach (Algorithm \ref{alg:2_1}) using BART within-type VIP as
the variable importance measure selects a predictor $x_{j}$ only if
\begin{equation}
\frac{1}{L}\sum\limits _{l=1}^{L}\mathbb{I}\left(\frac{\bar{c}_{L_{\text{rep}},j\cdot}}{c_{l,j\cdot}^{*}}>\frac{\bar{c}_{L_{\text{rep}},\text{type of }x_{j},\cdot}}{c_{l,\text{type of }x_{j},\cdot}^{*}}\right)\,>\,1-\alpha.\label{eq:2_7}
\end{equation}
For a binary predictor, the selection criteria (\ref{eq:2_7}) removes
the impact of continuous predictors, thereby not diluting the effect
of a relevant binary predictor. Figure \ref{fig:within-type-Bleich}
shows the variable selection result of this approach.
As opposed to the approach of  \citet{bleich2014variable}, 
both the two relevant binary predictors
$x_{1}$ and $x_{2}$ are clearly identified by this approach.

\subsection{BART Variable Importance Using Metropolis Ratios\label{subsec:3_2}}

While BART within-type VIP works well in including relevant predictors
in presence of a similar number of binary and continuous predictors,
it becomes problematic if the number of predictors of one type is
low. For example, when there are $99$ continuous predictors and $1$
binary predictor, the BART within-type VIP of the binary predictor
is always $1$ for both the original dataset and a null dataset. As
such, the binary predictor is never selected by the permutation-based
approach, whether it is relevant or not. Furthermore, when there are
more than two types of predictors, the computation of BART within-type
VIP becomes complicated. Hence, we propose a new type of variable
importance measure based on the Metropolis ratios calculated in the
Metropolis-Hastings steps for sampling new trees, which are not affected
by the issues above.

As mentioned in Section \ref{subsec:2_1}, a key feature of the
Metropolis-within-Gibbs sampler is the Bayesian backfitting procedure
for sampling $(T_{m},\boldsymbol{\mu}_{m})$, $1\leq m\leq M$, where
each $T_{m}$ is fit iteratively using the residual response $r_{-m}=y-\sum_{m^{\prime}\neq m}g(\boldsymbol{x};T_{m^{\prime}},\boldsymbol{\mu}_{m^{\prime}})$,
with all the other trees $T_{-m}=\{T_{m^{\prime}}\}_{m^{\prime}\neq m}$
held constant. Thus, each $(T_{m},\boldsymbol{\mu}_{m})$ can be obtained
in two sequential steps:
\begin{align*}
 & T_{m}\mid\boldsymbol{r}_{-m},\sigma^{2},\\
 & \boldsymbol{\mu}_{m}\mid T_{m},\boldsymbol{r}_{-m},\sigma^{2},
\end{align*}
where $\boldsymbol{r}_{-m}$ is the vector of residual responses.
The distribution of $T_{m}\mid\boldsymbol{r}_{-m},\sigma^{2}$ has
a closed form up to a normalizing constant and therefore can be sampled
by a Metropolis-Hastings algorithm. \citet{chipman1998bayesian} develop
a Metropolis-Hastings algorithm that proposes a new tree based on
the current tree using one of the four moves: BIRTH, DEATH, CHANGE
and SWAP. The BIRTH and DEATH proposals are the essential moves for
sufficient mixing of the Gibbs sampler, so both the CRAN \proglang{R} package \pkg{BART} (\citet{sparapani2021nonparametric})
and our package \pkg{BartMixVs} (\citet{luo2021package}) implement these two proposals,
each with equal probability. A BIRTH proposal turns a terminal node
into a splitting node and a DEATH proposal replaces a splitting node
leading to two terminal nodes by a terminal node. Thus, the proposed
tree $T^{*}$ is identical to the current tree $T_{m}$ except growing/pruning
one splitting node. 

We consider using the Metropolis ratio for accepting a BIRTH proposal
to construct a variable importance measure. For convenience, we suppress
the subscript $m$ in the following. The Metropolis ratio for accepting
a BIRTH proposal at a terminal node $\eta$ of the current tree $T$
can be written as
\begin{equation}
\pi_{\text{BIRTH}}\left(\eta\right) =  \min\left\{ 1, r(\eta)\right\},\label{eq:2_8}
\end{equation}
where
\begin{equation}
r(\eta) = \frac{P\left(T\mid T^{*}\right)P\left(T^{*}\mid\boldsymbol{r},\sigma^{2}\right)}{P\left(T^{*}\mid T\right)P\left(T\mid\boldsymbol{r},\sigma^{2}\right)}. \label{eq:2_9}
\end{equation}
The BART prior splits a node of depth $d$ into two child nodes of
$d+1$ with probability $\gamma(1+d)^{-\beta}$ for some $\gamma\in(0,1)$
and $\beta\in[0,\infty)$. Given that, the untruncated Metropolis ratio $r(\eta)$ can be
explicitly expressed as the product of three ratios:
\begin{eqnarray*}
\text{Nodes Ratio} & = & \frac{2b}{b+2}, \\
\text{Depth Ratio} & = & \frac{\gamma\left[1-\gamma / (2+d_{\eta})^{\beta} \right]^{2}}{(1+d_{\eta})^{\beta}-\gamma}, \\
\text{Likelihood Ratio} & = & \frac{P\left(\boldsymbol{r}\mid T^{*},\sigma^{2}\right)}{P\left(\boldsymbol{r}\mid T,\sigma^{2}\right)},
\end{eqnarray*}
where $b$ is the number of terminal nodes in the current tree $T$
and $d_{\eta}$ is the depth of the node $\eta$ in the current tree $T$.
The derivation can be found in Section S.2 of the
Supplementary Material.

Figure S.5 of the Supplementary Material 
shows the nodes ratios for different numbers of terminal nodes 
and the depth ratios for different depths when the hyper-parameters 
$\gamma$ and $\beta$ are set as default, i.e., $\gamma=0.95$ and $\beta=2$. 
From the figure, we can see that the product of nodes ratio and depth ratio is mostly affected by the
depth ratio which decreases as the proposed depth $d_{\eta}$ increases,
implying that $r(\eta)$ assigns a greater value to a shallower node
which typically contains more samples. Since the likelihood ratio
indicates the conditional improvement to fitting brought by the new
split, the untruncated Metropolis ratio $r(\eta)$ considers both
the proportion of samples affected at the node $\eta$ and the conditional
improvement to fitting brought by the new split at the node $\eta$.
However, it is not appropriate to directly use $r(\eta)$ 
to construct a variable importance measure, due to
the occurrence of extremely large $r(\eta)$ which dominates the others.
Instead, we use the Metropolis ratio in (\ref{eq:2_8}) to construct
the following BART Metropolis importance.

\begin{definition}
[BART Metropolis Importance\label{def:3_3}]Let $K$ be the number
of posterior samples obtained from a BART model. For every $k=1,\cdots,K$
and $j=1,\cdots,p$, define the average Metropolis acceptance ratio
per splitting rule using the predictor $x_{j}$ at the $k^{\text{th}}$
posterior sample as follows:
\begin{equation*}
\tilde{u}_{jk}\,=\,\frac{\sum\limits _{m=1}^{M}\sum\limits _{\eta\in\phi_{mk}}\left[\mathbb{I}\left(j_{\eta}=j\right)\pi_{\text{BIRTH}}\left(\eta\right)\right]}{\sum\limits _{m=1}^{M}\sum\limits _{\eta\in\phi_{mk}}\mathbb{I}\left(j_{\eta}=j\right)},
\end{equation*}
where $\phi_{mk}$ is the set of splitting nodes in the $k^{\text{th}}$
posterior sample of the $m^{\text{th}}$ tree and $j_{\eta}$ is the
indicator of the split variable at the node $\eta$. The BART Metropolis
Importance (MI) for the predictor $x_{j}$ is defined as the normalized
$\tilde{u}_{jk}$, averaged across $K$ posterior samples:
\begin{equation}
v_{j}^{\text{MI}}\,=\,\frac{1}{K}\sum\limits _{k=1}^{K}\frac{\tilde{u}_{jk}}{\sum\limits _{j=1}^{p}\tilde{u}_{jk}}.\label{eq:2_10}
\end{equation}
\end{definition}

\begin{figure}[ht]
\centering\includegraphics[width=0.65\columnwidth]{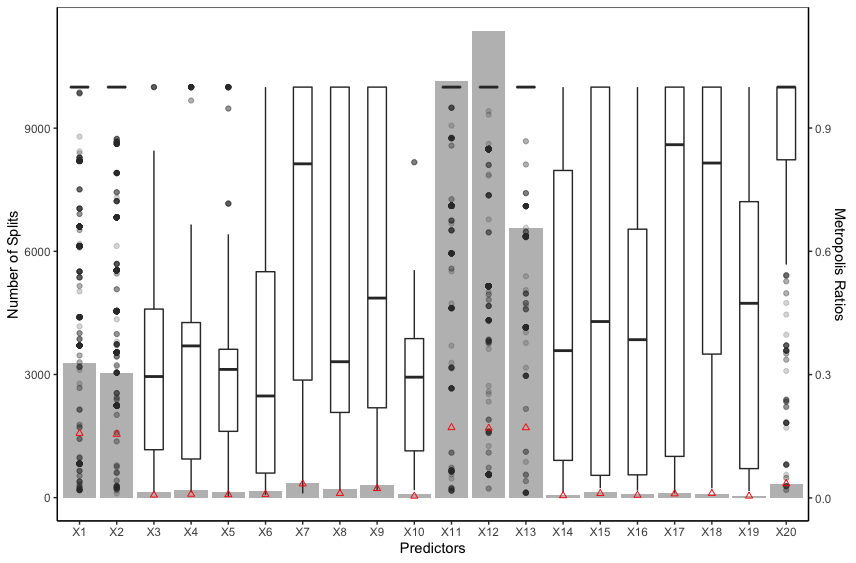}
\caption{\label{fig:Metropolis-ratio}
Each barplot (w.r.t. the left y-axis) depicts $c_{j\cdot}$, 
the number of splits using a predictor as the split variable
over all the posterior samples. 
Each boxplot (w.r.t. the right y-axis) depicts the Metropolis ratios $\pi_{\text{BIRTH}}(\eta)$
of the splits using a predictor as the split variable, over all the
posterior samples. Each red triangle (w.r.t. the right y-axis) displays
the BART MI $v_{j}^{\text{MI}}$ for a predictor. Posterior samples
are obtained from a BART model for Example \ref{exp:2}.}
\end{figure}

The intuition of BART MI is that a relevant predictor
is expected to be consistently accepted with a high Metropolis ratio.
Figure \ref{fig:Metropolis-ratio} shows that the Metropolis ratios
of the splits using relevant predictors ($x_{1}$, $x_{2}$, $x_{11}$,
$x_{12}$ and $x_{13}$) as the split variable have a higher median
and smaller standard error than those using irrelevant predictors,
implying that relevant predictors are consistently accepted with a
higher Metropolis ratio. 
BART MI and BART VIP have similar forms. 
The key difference is the ``kernel'' they use.
While BART VIP uses the number of splits $c_{jk}$ which tends to
be biased against binary predictors, BART MI uses the average Metropolis
acceptance ratio $\tilde{u}_{jk}$ which is similar between the relevant
binary predictors and relevant continuous predictors, as shown
in Figure \ref{fig:Metropolis-ratio}. Hence, the average Metropolis
acceptance ratio $\tilde{u}_{jk}$ not only helps BART MI distinguish
relevant and irrelevant predictors, but also helps BART MI
not be biased against predictors of a certain type. 

The vector of average Metropolis acceptance ratios $(\tilde{u}_{1k},\cdots,\tilde{u}_{pk})$
does not sum to $1$ and each of them varies from $0$ to $1$. We
normalize the vector $(\tilde{u}_{1k},\cdots,\tilde{u}_{pk})$ in
(\ref{eq:2_10}) based on the idea that $\tilde{u}_{jk}$'s are correlated
in the sense that once some predictors that explain the response more
are accepted with higher Metropolis ratios, the rest of predictors
will be accepted with lower Metropolis ratios or not be used. In
addition, we take the average over $K$ posterior samples in (\ref{eq:2_10})
because averaging further helps identify relevant predictors. In fact,
an irrelevant predictor can be accepted with a high Metropolis ratio
at some MCMC iterations, but it can be quickly removed from the model
by the DEATH proposal. As a result, the irrelevant predictor can have a
few large $\tilde{u}_{jk}$'s and many small $\tilde{u}_{jk}$'s,
thereby resulting in a small BART MI $v_{j}^{\text{MI}}$, as shown
in Figure \ref{fig:Metropolis-ratio}.

We further explore BART MI under null settings. 
We create a null dataset of Example \ref{exp:2} and fit a BART model on it.
Figure S.6 of the Supplementary Material
shows that not all the MIs, $v_{j}^{\text{MI}}$'s, converge to $1/p=0.05$,
implying that $1/p$ may not be a good selection threshold for $v_{j}^{\text{MI}}$'s.
We repeat the experiment of \citet{bleich2014variable} for BART MI
under null settings to explore the variation among $v_{j}^{\text{MI}}$'s.
Specifically, we create $100$ null datasets of the data generated from Example \ref{exp:2},  
and for each null dataset, we run BART $50$ times with different initial values of the hyper-parameters.
Let $v_{ijk}^{\text{MI}}$ be the BART MI for the predictor $x_j$
from the $k^{\text{th}}$ BART model on the $i^{\text{th}}$
null dataset. We investigate three nested variances listed in Table
\ref{tab:Three-nested-variance}.

\begin{table}[ht]
\caption{Three nested variances}
\label{tab:Three-nested-variance}
\centering%
\resizebox*{\textwidth}{!}{
\begin{tabular}{l>{\raggedright}p{12cm}}
\hline
Variance & Description \tabularnewline
\hline 
$s_{ij}^{2} = \frac{1}{49} \sum\limits _{k=1}^{50}(v_{ijk}^{\text{MI}}-\bar{v}_{ij\cdot}^{\text{MI}})^{2}$ 
& The variability of BART MI for the predictor $x_j$ in the $i^{\text{th}}$ dataset; 
$\bar{v}_{ij\cdot}^{\text{MI}}=(\sum_{k=1}^{50}v_{ijk}^{\text{MI}})/50$.\tabularnewline
$s_{j}^{2} = \frac{1}{99} \sum\limits_{i=1}^{100}(\bar{v}_{ij\cdot}^{\text{MI}}-\bar{v}_{\cdot j\cdot}^{\text{MI}})^{2}$ & The variability due to chance capitalization, i.e., fitting noise,
of BART for the predictor $x_j$ across datasets; $\bar{v}_{\cdot j\cdot}^{\text{MI}}=(\sum_{i,k=1}^{100,50}v_{ijk}^{\text{MI}})/(100\times50)$.\tabularnewline
$s^{2} = \frac{1}{19} \sum\limits_{j=1}^{20}(\bar{v}_{\cdot j\cdot}^{\text{MI}}-\bar{v}_{\cdot\cdot\cdot}^{\text{MI}})^{2}$ & The variability of BART MI across predictors; $\bar{v}_{\cdot\cdot\cdot}^{\text{MI}}=(\sum_{i,j,k=1}^{100,20,50}v_{ijk}^{\text{MI}})/(100\times20\times50)$.\tabularnewline
\hline 
\end{tabular}
}
\end{table}

\begin{figure}[ht]
\centering\includegraphics[width=0.7\columnwidth]{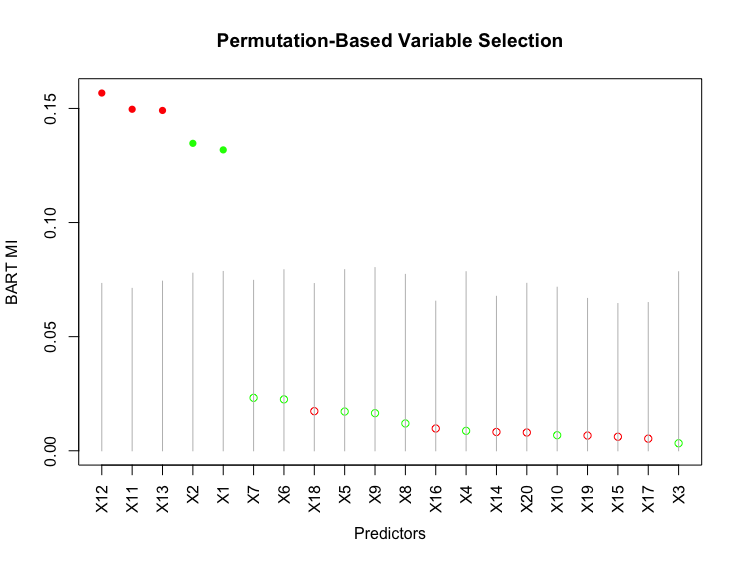}
\caption{\label{fig:MI-selection}Variable selection results of the permutation-based
approach using BART MI as the variable importance measure for Example \ref{exp:2}. 
The same setting and BART posterior samples as Figure S.1 of the Supplementary Material
and Figure \ref{fig:within-type-Bleich} are used. 
With BART MI, all the relevant predictors ($x_{1},$ $x_{2},$
$x_{11}$, $x_{12}$, and $x_{13}$) are identified.}
\end{figure}

Results of the experiment are shown in Figure S.7 of the Supplementary Material. 
The non-zero $s_{ij}$'s imply that there is variation in $v_{j}^{\text{MI}}$'s
among the repetitions of BART with different initial values, so it
is necessary to average MIs over a certain number of repetitions of
BART to get stable MIs. In practice, we find
that among the repetitions of BART, irrelevant
predictors occasionally get outliers of MI. Thus, instead of averaging
MIs over BART repetitions, we take the median of them. Similar to
\citet{bleich2014variable}, we also find that $10$ repetitions of
BART is sufficient to get stable MIs (see Figure S.8 of the Supplementary Material).
Figure S.7 also shows that the second type of standard deviation
$s_{j}$ is significantly greater than the median of $s_{ij}$'s for
each predictor, which suggests that BART tends to overfit noise. Since
the overall MI $\bar{v}_{\cdot\cdot\cdot}^{\text{MI}}$ is always
$1/p$, the relatively large $s$ indicates that not all the $\bar{v}_{\cdot j\cdot}^{\text{MI}}$
are approximately $1/p$, suggesting that it is not possible to determine
a single selection threshold for all the MIs. Therefore, we employ
the permutation-based approach (Algorithm \ref{alg:2_1}) to select
thresholds for each MI. A slight modification is applied to line 4
of Algorithm \ref{alg:2_1}: we take the median, rather than the average,
of BART MIs, over the repetitions of BART on the original dataset.
The variable selection result of this approach for Example \ref{exp:2} 
is shown in Figure \ref{fig:MI-selection}.

\subsection{Backward Selection with Two Filters\label{subsec:3_3}}

In this section, we propose an approach to select a subset of predictors
with which the BART model gives the best prediction. In general, to
select the best subset of predictors from the model space consisting
of $2^{p}$ possible models, one needs a search strategy over the
model space and decision-making rules to compare models. Here we use
the backward elimination approach to search the model space, mainly
because it is a deterministic approach and is efficient with a moderate
number of predictors. Typically, backward selection starts with the
full model with all the predictors, followed by comparing the deletion
of each predictor using a chosen model fit criterion and then deleting
the predictor whose loss gives the most statistically insignificant
deterioration of the model fit. This process is repeated until no
further predictors can be deleted without a statistically insignificant
loss of fit. Some popular model fit criterions such as AIC and BIC
are not available for BART, because the maximum likelihood estimates
are unavailable and the number of parameters in the model is hard
to determine. To overcome this issue, we propose a backward selection
approach with two easy-to-compute selection criterions as the decision-making
rules.

\begin{figure}[ht]
\centering\includegraphics[width=0.65\columnwidth]{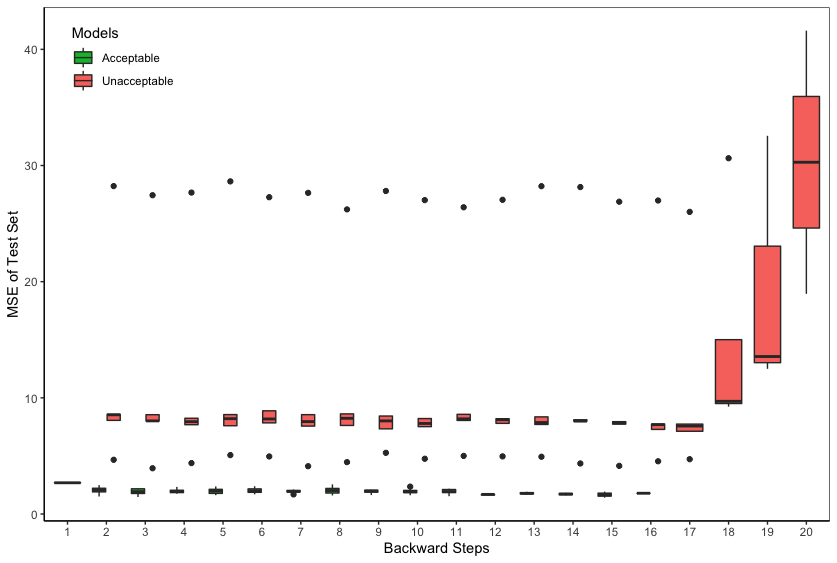}
\caption{\label{fig:mse}$\text{MSE}_{\text{test}}$'s of all the models evaluated
in the backward selection approach for Example \ref{exp:2}.
Each green (or red) boxplot depicts the $\text{MSE}_{\text{test}}$'s
of all the acceptable (or unacceptable) models evaluated at a backward
selection step. The $\text{MSE}_{\text{test}}$'s for acceptable models are close
to $0$ and have tiny standard error, so green boxplots are very narrow and 
near $0$.}
\end{figure}

We first split the dataset $\{y_{i},\boldsymbol{x}_{i}\}_{i=1}^{n}$
into a training set and a test set before starting the backward procedure.
To measure the model fit, we make use of the mean squared errors (MSE)
of the test set
\begin{equation*}
\text{MSE}_{\text{test}} = \frac{1}{n_{\text{test}}}\sum\limits _{i\in\text{test set}}\left[y_{i}-\hat{f}\left(\boldsymbol{x}_{i}\right)\right]^{2},
\end{equation*}
where $\hat{f}$ is the estimated sum-of-trees function. Figure \ref{fig:mse}
shows that $\text{MSE}_{\text{test}}$ can distinguish between acceptable
models and unacceptable models, where acceptable models are defined as those
models including all the relevant predictors and unacceptable models
are defined as those missing some relevant predictors. At each backward selection
step, we run BART models on the training set and compute the MSE of
the test set. The model with the smallest $\text{\ensuremath{\text{MSE}_{\text{test}}}}$
is chosen as the ``winner'' at that step. However, due to the lack
of stopping rules under such nonparametric setting, the backward selection
has to continue until there is only one predictor in the model, and
ultimately returns $p$ ``winner'' models with different model sizes
ranging from $1$ to $p$.

Given the $p$ ``winner'' models, we compare them using the expected
log pointwise predictive density based on leave-one-out (LOO) cross
validation:
\begin{equation}
\text{elpd}_{\text{loo}} = \sum\limits _{i=1}^{n}\log\left[f\left(y_{i}\mid\boldsymbol{y}_{-i}\right)\right]  =  \sum\limits _{i=1}^{n}\log\left[\int f\left(y_{i}\mid\theta\right)f\left(\theta\mid\boldsymbol{y}_{-i}\right)d\theta\right],\label{eq:2_11}
\end{equation}
where $\boldsymbol{y}_{-i}=\{y_{j}\}_{j\neq i}$, $\theta=\{\{T_{m},\boldsymbol{\mu}_{m}\}_{m=1}^{M},\sigma^{2}\}$
represents the set of BART parameters, $f\left(y_{i}\mid\boldsymbol{y}_{-i}\right)$
is the predictive density of $y_{i}$ given $\boldsymbol{y}_{-i}$,
$f(y_{i}\mid\theta)$ is the likelihood of $y_{i}$, and $f(\theta\mid\boldsymbol{y}_{-i})$
is the posterior density of BART given the observations $\boldsymbol{y}_{-i}$.
The quantity $\text{elpd}_{\text{loo}}$ not only measures the predictive
capability of a model, but also penalizes the complexity of the model,
i.e., the number of predictors used in the model. Hence, we choose
the model with the largest $\text{\ensuremath{\text{elpd}_{\text{loo}}}}$
as the best model. 

Direct computing (\ref{eq:2_11}) involves fitting BART $n$ times.
To avoid this, we use the approach of \citet{vehtari2017practical}
that estimates each $f(y_{i}\mid\boldsymbol{y}_{-i})$
using importance sampling with the full posterior distribution $f(\theta\mid\boldsymbol{y})$
as the sampling distribution and the smoothed ratios $\{f(\theta^{k}\mid\boldsymbol{y}_{-i}) / f(\theta^{k}\mid\boldsymbol{y}) \propto 1 / f(y_{i}\mid\theta^{k}) \}_{k=1}^{K}$
as the importance ratios, where $\{\theta^{k}\}_{k=1}^{K}$ are the
$K$ posterior samples from the full posterior distribution $f(\theta\mid\boldsymbol{y})$.
The smoothness is achieved by fitting a generalized Pareto distribution
to the upper tail of each set of the importance ratios $1 / f(y_{i}\mid\theta^{k}) \}_{k=1}^{K}$,
$i=1,\cdots,n$. Thus, $\text{elpd}_{\text{loo}}$ in (\ref{eq:2_11})
can be estimated by using the likelihoods $\{f(y_{i}\mid\theta^{k})=\phi(y_{i}\mid\hat{f}^{k}(\boldsymbol{x}_{i}),\sigma^{k})\}_{i,k=1}^{n,K}$
based on $K$ posterior samples $\{\hat{f}^{k}\}_{k=1}^{K}$ from
the BART model fit on the dataset $\{y_{i},\boldsymbol{x}_{i}\}_{i=1}^{n}$,
where $\hat{f}^{k}$ is the $k^{\text{th}}$ posterior sample of the
sum-of-trees function (\ref{eq:2_2}) and $\phi(\cdot\mid\mu,\sigma)$
is the normal density with mean $\mu$ and standard error $\sigma$.

The approach above can also be extended to the probit BART model
(\ref{eq:2_3}) by changing the $\text{MSE}_{\text{test}}$ to the
mean log loss (MLL) of the test set
\begin{equation}
\text{MLL}_{\text{test}} = -\frac{1}{n_{\text{test}}}\sum\limits _{i\in\text{test set}} \left[ y_{i}\log\left(\hat{p}_{i}\right)+  \left(1-y_{i}\right)\log\left(1-\hat{p}_{i}\right) \right]
\end{equation}
and replacing the normal likelihoods with the Bernoulli density $\{f(y_{i}\mid\theta^{k})=\hat{p}_{i}^{k}y_{i}+(1-\hat{p}_{i}^{k})(1-y_{i})\}_{i,k=1}^{n,K}$,
where $\hat{p}_{i}^{k}=\Phi(\hat{f}^{k}(x_{i}))$ is the estimated
probability based on the $k^{\text{th}}$ posterior sample of probit
BART for the $i^{\text{th}}$ observation and $\hat{p}_{i}=\Phi(\sum_{k=1}^{K}\hat{f}^{k}(x_{i}) / K)$.
We summarize this algorithm in Algorithm \ref{alg:2_2}. Although
this algorithm requires fitting BART $p(p-1)/2$ times, models at
the same backward selection step can be fitted in parallel, which
implies that the time complexity $O(p(p-1)/2)$ can be reduced to
$O(p)$ if there are sufficient computing resources.

\begin{algorithm}[ht]
\begin{algorithmic}[1]
\REQUIRE {$X$: predictors; $\boldsymbol{y}$: response; $s$: split ratio}
\ENSURE{ A subset of predictors.}
\STATE Randomly split the data $(\boldsymbol{y}, X)$ into $(\boldsymbol{y}_{\text{train}}, X_{\text{train}})$ and $(\boldsymbol{y}_{\text{test}}, X_{\text{test}})$ according to the split ratio $s$
\STATE Run BART, say $\text{Model}_1$, on $(\boldsymbol{y}_{\text{train}}, X_{\text{train}})$ with all the predictors and compute the LOO score: $\text{elpd}_{\text{loo}}^1$
\FOR {$l=1,\cdots,p-1$}
\FOR {$t = 1,\cdots,p-l+1$}
\STATE Remove the $t^{\text{th}}$ predictor from the predictors used in $\text{Model}_l$
\STATE Run BART, say $\text{Model}_{l+1,t}$, with the remaining predictors on $(\boldsymbol{y}_{\text{train}}, X_{\text{train}})$
\STATE Compute the MSE of the test set: $\text{MSE}_{\text{test}}^{l+1,t}$ (or $\text{MLL}_{\text{test}}^{l+1,t}$ if $\boldsymbol{y}$ is binary)
\STATE Compute the LOO score of the training set: $\text{elpd}_{\text{loo}}^{l+1,t}$
\ENDFOR
\STATE Find $t^*$ that minimizes $\{\text{MSE}_{\text{test}}^{l+1,t}\}_{t=1}^{p-l+1}$ (or $\{\text{MLL}_{\text{test}}^{l+1,t}\}_{t=1}^{p-l+1}$ if $\boldsymbol{y}$ is binary)
\STATE Denote $\text{Model}_{l+1,t^*}$ by $\text{Model}_{l+1}$ and $\text{elpd}_{\text{loo}}^{l+1,t^*}$ by $\text{elpd}_{\text{loo}}^{l+1} $
\ENDFOR
\STATE Find $l^*$ that maximizes $\{\text{elpd}_{\text{loo}}^{l}\}_{l=1}^{p}$;
\STATE Return $\text{Model}_{l^*}$.
\end{algorithmic} 
\caption{Backward selection with two filters}\label{alg:2_2}
\end{algorithm}

\section{Simulations\label{sec:4}}

In this section, we compare the proposed variable selection approaches:
(1) permutation-based approach using within-type BART VIP, (2) permutation-based
approach using BART MI, and (3) backward selection with two filters,
with the existing BART-based variable selection approaches: (4) median
probability model from DART, (5) permutation-based approach using
BART VIP, and (6) ABC Bayesian forest. For the permutation-based approaches
(1), (2) and (5), we set $L_{\text{rep}}=10$, $L=100$ and $\alpha=0.05$
(see Algorithm \ref{alg:2_1}), and use $M=20$ trees for BART to
encourage competition among the predictors. The three permutation-based
approaches are based on the same posterior samples and the one using
within-type BART VIP is only applied to scenarios with mixed-type
predictors. For the backward selection procedure (3), we set $s=80\%$
(see Algorithm \ref{alg:2_2}) and use $M=50$ trees for BART to ensure
its prediction power. For DART (4), we consider $M=20$ and $M=200$
trees, respectively. For ABC Bayesian forest, as \citet{liu2019variable}
suggest, we assume the prior $\pi(\mathcal{S})$ on the model space
to be the beta-binomial prior with $\theta\sim\mathrm{Beta}(1,1)$,
set the split ratio as $50\%$, and use $M=10$ and $M=20$ trees
for BART respectively; we generate $1000$ ABC samples, rank the samples
by MSE in ascending order if the response variable is continuous (or
by MLL if the response variable is binary), and keep the top $10\%$
of samples for selection; we report selection results for two selection
thresholds $0.25$ and $0.5$. All the methods above, except ABC Bayesian
forest, burn in the first 1000 and keep 1000 posterior samples in
each BART run. ABC Bayesian forest burns in $200$ posterior samples
and saves the $201^{\text{th}}$ posterior sample, as \citet{liu2019variable}
suggest. Other hyper-parameters of BART are the same as the default
recommended by \citet{chipman2010bart}. The simulation results were
obtained using the \proglang{R} package \pkg{BartMixVs} (\citet{luo2021package})
which is briefly introduced
in Section S.3 of the Supplementary Material.

To evaluate variable selection results, we consider precision (prec),
recall (rec), $F_{1}$ score, and the proportion of the replications
that miss at least one relevant predictor ($r_{\text{miss}}$), given
by
\begin{align*}
& \text{prec} =  \frac{\text{TP}}{\text{TP}+\text{FP}},\,\text{rec} = \frac{\text{TP}}{\text{TP}+\text{FN}},\, F_{1}=\frac{2\cdot\text{prec}\cdot\text{rec}}{\text{prec}+\text{rec}}, \\
& r_{\text{miss}} = \frac{\text{\# of replications missing at least one relevant predictor}}{\text{\# of total replications}},
\end{align*}
where $\text{TP}$ is the number of relevant predictors correctly
selected, $\text{FP}$ is the number of predictors incorrectly selected
and $\text{FN}$ is the number of relevant predictors not selected.
High precision indicates good ability of excluding irrelevant predictors
and high recall indicates good ability of including relevant predictors.
The $F_{1}$ score is a balanced score between precision and recall.
The lowest $r_{\text{miss}}$ score is $0$, implying that the method
does not miss any relevant predictors over all the replications.

We evaluate the aforementioned variable selection approaches under
four possible combination scenarios of 
$\left\{ \text{continuous response }y,\text{ binary response }y\right\} \times 
\left\{ \text{continuous predictors }\boldsymbol{x},\text{ mixed-type predictors }\boldsymbol{x}\right\}$.
Each scenario is replicated 100 times. The simulation results for the
scenario of binary response and continuous predictors can be found
in Section S.4.2 of the Supplementary Material.

\subsection{Continuous Response and Continuous Predictors}

In this setting, we consider two scenarios. Scenario C.C.1
is an example from \citet{friedman1991multivariate}; Scenario
C.C.2, which includes correlation between predictors, is
borrowed from \citet{zhu2015reinforcement}. For both of them, we
consider $n=500$, $p\in\{50,200\}$ and $\sigma^{2}=1$.

\begin{scenarioCC1}
Sample the predictors $x_{1},\cdots,x_{p}$ from
$\text{Uniform}(0,1)$ independently and the response
$y$ from $\text{Normal}(f_{0}(\boldsymbol{x}),\sigma^{2})$,
where
\begin{equation}
f_{0}\left(\boldsymbol{x}\right) = 10\sin\left(\pi x_{1}x_{2}\right)+20\left(x_{3}-0.5\right)^{2}+10x_{4}+5x_{5}.\label{eq:2_12}
\end{equation}
\end{scenarioCC1}

\begin{scenarioCC2}
Sample the predictors $x_{1},\cdots,x_{p}$ from
$\text{Normal}(0,\Sigma)$ with $\Sigma_{jk}=0.3^{|j-k|}$, $j,k=1,\cdots,p$,
and the response $y$ from $\text{Normal}(f_{0}(\boldsymbol{x}),\sigma^{2})$,
where
\begin{equation}
f_{0}\left(\boldsymbol{x}\right) = 2x_{1}x_{4}+2x_{7}x_{10}.\label{eq:2_13}
\end{equation}
\end{scenarioCC2}

Results are given in Table \ref{tab:cYcX}. When the predictors are
independent (Scenario C.C.1) and $p=50$, none of the
methods miss any relevant predictors, i.e., $r_{\text{miss}}=0$ and
$\text{rec}\equiv1$. The permutation-based approach using BART
VIP and the ABC Bayesian forest methods, except the one using $M=20$
trees and selection threshold $0.25$ (i.e., ABC-20-0.25), are the
best in terms of precision and the $F_{1}$ scores, followed by the
permutation-based approach using BART MI and DART with $M=20$ trees.
When $p=200$, only the ABC Bayesian forest with $M=10$ trees and
selection threshold $0.5$ (i.e., ABC-10-0.50) fails to include all
the relevant predictors. As discussed in Section \ref{subsec:2_2_4},
ABC Bayesian forest using a small number of trees and keeping a small
number of posterior samples makes it easy to miss relevant predictors,
especially when the number of predictors is large and the selection
threshold is high. As shown in Table \ref{tab:cYcX}, increasing the
number of trees to $M=20$ or decreasing the selection threshold to
$0.25$ can improve the result. The other three ABC Bayesian forest
methods achieve the best precision and $F_{1}$ scores, followed by
the two permutation-based approaches. The backward selection procedure
is not a top choice in terms of precision, but its precision ($0.75$
for $p=50$ and $0.74$ for $p=200$) is acceptable in the sense that
it only includes about $1.7$ irrelevant predictors on average. In
fact, we find that the LOO score used in the backward selection, though
penalizing the model complexity, appears to have difficulty in distinguishing
between the true model and an acceptable model with a similar number
of predictors, thereby resulting in a relatively low precision score.

\begin{table}[ht]
\caption{Simulation results for the scenarios using continuous
$y$ and continuous $\boldsymbol{x}$. Scenario C.C.1 is Friedman's
example; Scenario C.C.2 includes correlation between predictors.
Each score is the average across $100$ replications and Monte Carlo
standard error is given inside the parentheses.}
\label{tab:cYcX}
\resizebox*{\textwidth}{!}{
\begin{tabular}{lccccccccc}
\hline 
 & \multicolumn{9}{c}{Scenario C.C.1}\tabularnewline
\cline{2-10} \cline{3-10} \cline{4-10} \cline{5-10} \cline{6-10} \cline{7-10} \cline{8-10} \cline{9-10} \cline{10-10} 
 & \multicolumn{4}{c}{\textit{$n=500,p=50,\sigma^{2}=1$}} &  & \multicolumn{4}{c}{\textit{$n=500,p=200,\sigma^{2}=1$}}\tabularnewline
\cline{2-5} \cline{3-5} \cline{4-5} \cline{5-5} \cline{7-10} \cline{8-10} \cline{9-10} \cline{10-10} 
 & $r_{\text{miss}}$ & rec & prec & $F_{1}$ &  & $r_{\text{miss}}$ & rec & prec & $F_{1}$\tabularnewline
\hline 
DART-20 & 0 & 1(0) & 0.87(0.01) & 0.93(0.01) &  & 0 & 1(0) & 0.86(0.01) & 0.92(0.01)\tabularnewline
DART-200 & 0 & 1(0) & 0.7(0.02) & 0.81(0.01) &  & 0 & 1(0) & 0.63(0.02) & 0.76(0.01)\tabularnewline
ABC-10-0.25 & 0 & 1(0) & 1(0) & 1(0) &  & 0 & 1(0) & 1(0) & 1(0)\tabularnewline
ABC-10-0.50 & 0 & 1(0) & 1(0) & 1(0) &  & 0.33 & 0.93(0.01) & 1(0) & 0.96(0.01)\tabularnewline
ABC-20-0.25 & 0 & 1(0) & 0.5(0.01) & 0.66(0.01) &  & 0 & 1(0) & 1(0) & 1(0)\tabularnewline
ABC-20-0.50 & 0 & 1(0) & 1(0) & 1(0) &  & 0 & 1(0) & 1(0) & 1(0)\tabularnewline
Permute (VIP) & 0 & 1(0) & 1(0) & 1(0) &  & 0 & 1(0) & 0.95(0.01) & 0.97(0)\tabularnewline
Permute (MI) & 0 & 1(0) & 0.98(0.01) & 0.99(0) &  & 0 & 1(0) & 0.94(0.01) & 0.97(0.01)\tabularnewline
Backward & 0 & 1(0) & 0.75(0.02) & 0.83(0.02) &  & 0 & 1(0) & 0.74(0.02) & 0.83(0.02)\tabularnewline
\hline 
 & \multicolumn{9}{c}{Scenario C.C.2}\tabularnewline
\cline{2-10} \cline{3-10} \cline{4-10} \cline{5-10} \cline{6-10} \cline{7-10} \cline{8-10} \cline{9-10} \cline{10-10} 
 & \multicolumn{4}{c}{\textit{$n=500,p=50,\sigma^{2}=1$}} &  & \multicolumn{4}{c}{\textit{$n=500,p=200,\sigma^{2}=1$}}\tabularnewline
\cline{2-5} \cline{3-5} \cline{4-5} \cline{5-5} \cline{7-10} \cline{8-10} \cline{9-10} \cline{10-10} 
 & $r_{\text{miss}}$ & rec & prec & $F_{1}$ &  & $r_{\text{miss}}$ & rec & prec & $F_{1}$\tabularnewline
\hline 
DART-20 & 0.43 & 0.8(0.03) & 0.86(0.02) & 0.8(0.02) &  & 0.97 & 0.37(0.03) & 0.47(0.04) & 0.39(0.03)\tabularnewline
DART-200 & 0 & 1(0) & 0.83(0.02) & 0.9(0.01) &  & 0.64 & 0.65(0.03) & 0.49(0.03) & 0.53(0.03)\tabularnewline
ABC-10-0.25 & 0.02 & 1(0) & 0.69(0.02) & 0.8(0.01) &  & 1 & 0.08(0.01) & 0.28(0.04) & 0.13(0.02)\tabularnewline
ABC-10-0.50 & 0.82 & 0.61(0.02) & 0.98(0.01) & 0.73(0.02) &  & 1 & 0(0) & 0.02(0.01) & 0.01(0.01)\tabularnewline
ABC-20-0.25 & 0 & 1(0) & 0.12(0) & 0.22(0) &  & 0.99 & 0.28(0.02) & 0.45(0.04) & 0.33(0.03)\tabularnewline
ABC-20-0.50 & 0.1 & 0.97(0.01) & 0.97(0.01) & 0.97(0.01) &  & 1 & 0.01(0) & 0.04(0.02) & 0.02(0.01)\tabularnewline
Permute (VIP) & 0 & 1(0) & 0.98(0.01) & 0.99(0) &  & 0.71 & 0.68(0.03) & 0.26(0.01) & 0.37(0.01)\tabularnewline
Permute (MI) & 0 & 1(0) & 0.89(0.01) & 0.93(0.01) &  & 0.93 & 0.43(0.03) & 0.27(0.02) & 0.32(0.02)\tabularnewline
Backward & 0 & 1(0) & 0.86(0.02) & 0.92(0.01) &  & 0.02 & 0.99(0.01) & 0.79(0.03) & 0.84(0.03)\tabularnewline
\hline 
\end{tabular}
}
\end{table}

When the predictors are correlated (Scenario C.C.2) and $p=50$,
the two permutation-based approaches, the backward selection approach,
DART with $M=200$ trees, and the ABC-20-0.25 approach successfully
identify all the relevant predictors for all the replications; the
first four of them have competitive precision and $F_{1}$ scores.
When more correlated and irrelevant predictors are included $(p=200)$,
the backward selection procedure has superior performance in all metrics.
It identifies all the relevant predictors for $98\%$ of the times
and achieves the highest precision ($0.79$) and $F_{1}$ scores ($0.84$).
Compared with the independent setting (Scenario C.C.1), ABC
Bayesian forest suffers the most from the multicollinearity as we found that 
this approach often does not select out any predictors.

With the first three examples in Table \ref{tab:cYcX} being considered,
the two proposed approaches along with the approach of \citet{bleich2014variable}
and DART with $M=200$ trees consistently perform well in identifying
all the relevant predictors. For Scenario C.C.2 with $p=200$
predictors, the backward selection approach significantly outperforms
other methods.

\subsection{Continuous Response and Mixed-Type Predictors}

In this setting, we consider Scenario C.M.1, a modified
Friedman's example, and Scenario C.M.2,
an example including correlation between predictors. We consider
$n\in\{500,1000\}$, $p\in\{50,200\}$ and $\sigma^{2}\in\{1,10\}$
for Scenario C.M.1, and $n\in\{500,1000\}$ and
$\sigma^{2}\in\{1,10\}$ for Scenario C.M.2.

\begin{scenarioCM1}
Sample the predictors $x_{1},\cdots,x_{\left\lceil p/2\right\rceil }$
from $\text{Bernoulli}(0.5)$ independently, $x_{\left\lceil p/2\right\rceil +1},\cdots,x_{p}$ from $\text{Uniform}(0,1)$ independently, and the response $y$ from $\text{Normal}(f_{0}(\boldsymbol{x}),\sigma^{2})$,
where
\begin{equation}
f_{0}\left(\boldsymbol{x}\right) = 10\sin\left(\pi x_{\left\lceil p/2 \right\rceil +1}x_{\left\lceil p/2 \right\rceil +2}\right) + 20\left(x_{\left\lceil p/2 \right\rceil +3}-0.5\right)^{2}+10x_{1}+5x_{2}. \label{eq:2_14}
\end{equation}
\end{scenarioCM1}

\begin{scenarioCM2}
Sample the predictors $x_{1},\cdots,x_{20}$ from
$\text{Bernoulli}(0.2)$ independently, $x_{21},\cdots,x_{40}$ from
$\text{Bernoulli(0.5)}$ independently, $x_{41},\cdots,x_{84}$ from
a multivariate normal distribution with mean $0$, variance $1$ and
correlation $0.3$, and the response $y$ from $\text{Normal}(f_{0}(\boldsymbol{x}),\sigma^{2})$,
where
\begin{equation}
f_{0}\left(\boldsymbol{x}\right) = -4+x_{1}+\sin\left(\pi x_{1}x_{44}\right)-x_{21}+0.6x_{41}x_{42}- \exp\left[-2\left(x_{42}+1\right)^{2}\right]-x_{43}^{2}+0.5x_{44}.\label{eq:2_15} 
\end{equation}
\end{scenarioCM2}

We report the simulation results for Scenario C.M.1 with the settings, $n=500$,
$p \in \{50, 200\}$ and $\sigma^2=1$, and Scenario C.M.2 with the settings, 
$n \in \{500, 1000\}$ and $\sigma^2=1$, in Table \ref{tab:cYmX}. Simulation 
results for other settings are reported in Table S.1 of the Supplementary
Material.

\begin{table}[ht]
\caption{A part of simulation results for Scenario C.M.1 where predictors are of mixed type and independent and Scenario C.M.2 where predictors are of mixed type and correlated. The rest of simulation results can be found in Table S.1 of the Supplementary Material.}
\label{tab:cYmX}
\resizebox*{\textwidth}{!}{
\begin{tabular}{lccccccccc}
\hline 
 & \multicolumn{9}{c}{Scenario C.M.1}\tabularnewline
\cline{2-10} \cline{3-10} \cline{4-10} \cline{5-10} \cline{6-10} \cline{7-10} \cline{8-10} \cline{9-10} \cline{10-10} 
 & \multicolumn{4}{c}{\textit{$n=500,p=50,\sigma^{2}=1$}} &  & \multicolumn{4}{c}{\textit{$n=500,p=200,\sigma^{2}=1$}}\tabularnewline
\cline{2-5} \cline{3-5} \cline{4-5} \cline{5-5} \cline{7-10} \cline{8-10} \cline{9-10} \cline{10-10} 
 & $r_{\text{miss}}$ & rec & prec & $F_{1}$ &  & $r_{\text{miss}}$ & rec & prec & $F_{1}$\tabularnewline
\hline 
DART-20 & 0 & 1(0) & 0.96(0.01) & 0.98(0) &  & 0 & 1(0) & 0.95(0.01) & 0.97(0)\tabularnewline
DART-200 & 0 & 1(0) & 0.65(0.02) & 0.78(0.01) &  & 0 & 1(0) & 0.63(0.02) & 0.76(0.01)\tabularnewline
ABC-10-0.25 & 0 & 1(0) & 0.99(0) & 0.99(0) &  & 0 & 1(0) & 1(0) & 1(0)\tabularnewline
ABC-10-0.50 & 0 & 1(0) & 1(0) & 1(0) &  & 0.56 & 0.89(0.01) & 1(0) & 0.94(0.01)\tabularnewline
ABC-20-0.25 & 0 & 1(0) & 0.62(0.01) & 0.76(0.01) &  & 0 & 1(0) & 0.95(0.01) & 0.97(0)\tabularnewline
ABC-20-0.50 & 0 & 1(0) & 1(0) & 1(0) &  & 0 & 1(0) & 1(0) & 1(0)\tabularnewline
Permute (VIP) & 0.29 & 0.94(0.01) & 1(0) & 0.97(0.01) &  & 0 & 1(0) & 0.85(0.01) & 0.91(0.01)\tabularnewline
Permute (Within-Type VIP) & 0 & 1(0) & 0.99(0.01) & 0.99(0) &  & 0 & 1(0) & 0.73(0.02) & 0.84(0.01)\tabularnewline
Permute (MI) & 0 & 1(0) & 0.97(0.01) & 0.98(0) &  & 0 & 1(0) & 0.82(0.01) & 0.9(0.01)\tabularnewline
Backward & 0 & 1(0) & 0.76(0.02) & 0.84(0.02) &  & 0 & 1(0) & 0.82(0.02) & 0.88(0.02)\tabularnewline
\hline 
 & \multicolumn{9}{c}{Scenario C.M.2}\tabularnewline
\cline{2-10} \cline{3-10} \cline{4-10} \cline{5-10} \cline{6-10} \cline{7-10} \cline{8-10} \cline{9-10} \cline{10-10} 
 & \multicolumn{4}{c}{$n=500,p=84,\sigma^{2}=1$} &  & \multicolumn{4}{c}{$n=1000,p=84,\sigma^{2}=1$}\tabularnewline
\cline{2-5} \cline{3-5} \cline{4-5} \cline{5-5} \cline{7-10} \cline{8-10} \cline{9-10} \cline{10-10} 
 & $r_{\text{miss}}$ & rec & prec & $F_{1}$ &  & $r_{\text{miss}}$ & rec & prec & $F_{1}$\tabularnewline
\hline 
DART-20 & 0.37 & 0.94(0.01) & 0.91(0.01) & 0.92(0.01) &  & 0.09 & 0.98(0) & 0.91(0.01) & 0.94(0.01)\tabularnewline
DART-200 & 0.17 & 0.97(0.01) & 0.49(0.01) & 0.64(0.01) &  & 0.01 & 1(0) & 0.58(0.01) & 0.73(0.01)\tabularnewline
ABC-10-0.25 & 0.95 & 0.81(0.01) & 0.94(0.01) & 0.87(0.01) &  & 0.54 & 0.91(0.01) & 0.99(0) & 0.94(0)\tabularnewline
ABC-10-0.50 & 1 & 0.69(0.01) & 1(0) & 0.81(0.01) &  & 0.96 & 0.82(0.01) & 1(0) & 0.9(0)\tabularnewline
ABC-20-0.25 & 0.57 & 0.9(0.01) & 0.44(0.01) & 0.58(0.01) &  & 0.05 & 0.99(0) & 0.69(0.01) & 0.8(0.01)\tabularnewline
ABC-20-0.50 & 1 & 0.78(0.01) & 0.99(0) & 0.87(0.01) &  & 0.7 & 0.88(0.01) & 1(0) & 0.93(0)\tabularnewline
Permute (VIP) & 0.18 & 0.97(0.01) & 0.93(0.01) & 0.95(0.01) &  & 0.51 & 0.9(0.01) & 0.96(0.01) & 0.93(0.01)\tabularnewline
Permute (Within-Type VIP) & 0.44 & 0.93(0.01) & 0.86(0.02) & 0.89(0.01) &  & 0.06 & 0.99(0.01) & 0.92(0.01) & 0.95(0.01)\tabularnewline
Permute (MI) & 0.07 & 0.99(0) & 0.85(0.01) & 0.91(0.01) &  & 0 & 1(0) & 0.89(0.01) & 0.94(0.01)\tabularnewline
Backward & 0.11 & 0.98(0.01) & 0.47(0.03) & 0.57(0.03) &  & 0 & 1(0) & 0.75(0.02) & 0.83(0.02)\tabularnewline
\hline 
\end{tabular}
}
\end{table}

When the predictors are independent (Scenario C.M.1) and 
$p=50$, all the methods, except the permutation-based approach using
BART VIP, are able to identify all the relevant predictors. The two
proposed permutation-based approaches, DART with $M=20$ trees, and
the ABC Bayesian forest approaches except ABC-20-0.25 have comparable
best precision and $F_{1}$ scores, followed by the backward selection
procedure. When the dimension is increased to $p=200$, the ABC-10-0.50
method again fails to include all the relevant predictors, like Scenario
C.C.1. The permutation-based approach using BART VIP does not perform
poorly in this case, because as $p$ gets larger, $\bar{c}_{L_{\text{rep}},j\cdot}$
stays in a similar magnitude and $c_{l,j\cdot}^{*}$ gets smaller
(see the left subfigure of Figure S.4 of the Supplementary
Material). As such, the
offset effect discussed in Section \ref{subsec:3_1} disappears.
The three proposed approaches are still able to identify all the relevant
predictors, but their precision scores are slightly worse than other
methods.

When the mixed-type predictors are correlated (Scenario C.M.2),
variable selection becomes more challenging. When $\sigma^{2}=1$,
the permutation-based approach using BART MI and the backward selection
procedure are the best in terms of the $r_{\text{miss}}$ and recall
scores. Furthermore, only these two methods successfully identify
all the relevant predictors over all the replications when $n=1000$.
When high noise $(\sigma^{2}=10)$ is included, all the methods have
difficulty identifying all the true predictors. 

In terms of of the recall and $r_{\text{miss}}$ scores, the permutation-based
approach using BART MI and the backward selection approach are always
the top choices for all the examples in Table \ref{tab:cYmX} and
Table S.1, except Scenario C.M.2 with $\sigma^{2}=10$.
Compared to the other two proposed approaches, the permutation-based
approach using BART within-type VIP suffers from multicollinearity,
though it does improve the approach of \citet{bleich2014variable}
in presence of a small number of mixed-type predictors.

\subsection{Binary Response and Mixed-Type Predictors}

In this setting, we consider two scenarios, Scenario B.M.1
and Scenario B.M.2. For Scenario B.M.1,
we consider $n\in\{500,1000\}$ and $p\in\{50,200\}$; for Scenario B.M.2, we consider $n\in\{500,1000\}$.

\begin{scenarioBM1}
Sample the predictors $x_{1},\cdots,x_{\left\lceil p/2\right\rceil }$
from $\text{Bernoulli}(0.5)$ independently, $x_{\left\lceil p/2\right\rceil +1},\cdots,x_{p}$
from $\text{Uniform}(0,1)$ independently, and the response $y$ from $\text{Bernoulli}(\Phi(f_{0}(\boldsymbol{x}))$,
where $f_{0}(\boldsymbol{x})$ is defined in (\ref{eq:2_14}).
\end{scenarioBM1}

\begin{scenarioBM2}
Sample the predictors $x_{1},\cdots,x_{20}$ from
$\text{Bernoulli}(0.2)$ independently, $x_{21},\cdots,x_{40}$ from
$\text{Bernoulli(0.5)}$ independently, $x_{41},\cdots,x_{84}$ from
a multivariate normal distribution with mean $0$, variance $1$ and
correlation $0.3$, and the response $y$ from $\text{Bernoulli}(\Phi(f_{0}(\boldsymbol{x}))$,
where $f_{0}(\boldsymbol{x})$ is defined in (\ref{eq:2_15}).
\end{scenarioBM2}

\begin{table}[ht]
\caption{Simulation results for the scenarios using binary
response and mixed-type predictors. Predictors in Scenario
B.M.1 are independent and predictors in Scenario B.M.2 are
correlated.}
\label{tab:bYmX}
\resizebox*{\textwidth}{!}{
\begin{tabular}{lccccccccc}
\hline 
 & \multicolumn{9}{c}{\textit{Scenario B.M.1}}\tabularnewline
\cline{2-10} \cline{3-10} \cline{4-10} \cline{5-10} \cline{6-10} \cline{7-10} \cline{8-10} \cline{9-10} \cline{10-10} 
 & \multicolumn{4}{c}{\textit{$n=500,p=50$}} &  & \multicolumn{4}{c}{\textit{$n=500,p=200$}}\tabularnewline
\cline{2-5} \cline{3-5} \cline{4-5} \cline{5-5} \cline{7-10} \cline{8-10} \cline{9-10} \cline{10-10} 
 & $r_{\text{miss}}$ & rec & prec& $F_{1}$ &  & $r_{\text{miss}}$ & rec & prec & $F_{1}$\tabularnewline
\hline 
DART-20 & 0.26 & 0.95(0.01) & 0.99(0.01) & 0.96(0.01) &  & 0.65 & 0.87(0.01) & 0.99(0) & 0.92(0.01)\tabularnewline
DART-200 & 0.13 & 0.97(0.01) & 0.75(0.02) & 0.83(0.01) &  & 0.53 & 0.89(0.01) & 0.79(0.02) & 0.83(0.01)\tabularnewline
ABC-10-0.25 & 0.55 & 0.89(0.01) & 0.87(0.01) & 0.87(0.01) &  & 1 & 0.8(0) & 0.98(0.01) & 0.88(0)\tabularnewline
ABC-10-0.50 & 0.97 & 0.81(0) & 1(0) & 0.89(0) &  & 1 & 0.69(0.01) & 1(0) & 0.81(0.01)\tabularnewline
ABC-20-0.25 & 0.03 & 0.99(0) & 0.18(0) & 0.31(0) &  & 0.92 & 0.82(0.01) & 0.85(0.01) & 0.83(0.01)\tabularnewline
ABC-20-0.50 & 0.85 & 0.83(0.01) & 0.98(0.01) & 0.9(0) &  & 1 & 0.78(0.01) & 0.99(0) & 0.87(0)\tabularnewline
Permute (VIP) & 0.1 & 0.98(0.01) & 0.99(0) & 0.99(0) &  & 0.16 & 0.97(0.01) & 0.81(0.01) & 0.88(0.01)\tabularnewline
Permute (Within-Type VIP) & 0.08 & 0.98(0.01) & 1(0) & 0.99(0) &  & 0.22 & 0.96(0.01) & 0.8(0.02) & 0.87(0.01)\tabularnewline
Permute (MI) & 0.1 & 0.98(0.01) & 0.96(0.01) & 0.97(0.01) &  & 0.25 & 0.95(0.01) & 0.79(0.01) & 0.86(0.01)\tabularnewline
Backward & 0.09 & 0.98(0.01) & 0.93(0.02) & 0.94(0.01) &  & 0.37 & 0.93(0.01) & 0.87(0.02) & 0.88(0.01)\tabularnewline
\hline 
 & \multicolumn{4}{c}{$n=1000,p=50$} &  & \multicolumn{4}{c}{$n=1000,p=200$}\tabularnewline
\cline{2-5} \cline{3-5} \cline{4-5} \cline{5-5} \cline{7-10} \cline{8-10} \cline{9-10} \cline{10-10} 
 & $r_{\text{miss}}$ & rec & prec & $F_{1}$ &  & $r_{\text{miss}}$ & rec & prec & $F_{1}$\tabularnewline
\hline 
DART-20 & 0.01 & 1(0) & 0.98(0.01) & 0.99(0) &  & 0.17 & 0.97(0.01) & 0.98(0.01) & 0.97(0.01)\tabularnewline
DART-200 & 0 & 1(0) & 0.95(0.01) & 0.97(0.01) &  & 0.03 & 0.99(0) & 0.95(0.01) & 0.97(0.01)\tabularnewline
ABC-10-0.25 & 0.01 & 1(0) & 0.95(0.01) & 0.97(0.01) &  & 0.82 & 0.84(0.01) & 0.99(0) & 0.9(0)\tabularnewline
ABC-10-0.50 & 0.53 & 0.89(0.01) & 1(0) & 0.94(0.01) &  & 1 & 0.8(0) & 1(0) & 0.89(0)\tabularnewline
ABC-20-0.25 & 0 & 1(0) & 0.24(0) & 0.39(0) &  & 0.22 & 0.96(0.01) & 0.91(0.01) & 0.93(0.01)\tabularnewline
ABC-20-0.50 & 0.02 & 1(0) & 0.99(0) & 0.99(0) &  & 0.96 & 0.81(0) & 1(0) & 0.89(0)\tabularnewline
Permute (VIP) & 0 & 1(0) & 1(0) & 1(0) &  & 0 & 1(0) & 0.85(0.01) & 0.92(0.01)\tabularnewline
Permute (Within-Type VIP) & 0 & 1(0) & 0.99(0) & 1(0) &  & 0 & 1(0) & 0.85(0.02) & 0.91(0.01)\tabularnewline
Permute (MI) & 0 & 1(0) & 0.96(0.01) & 0.98(0) &  & 0 & 1(0) & 0.82(0.01) & 0.89(0.01)\tabularnewline
Backward & 0 & 1(0) & 0.91(0.01) & 0.95(0.01) &  & 0 & 1(0) & 0.9(0.01) & 0.94(0.01)\tabularnewline
\hline 
 & \multicolumn{9}{c}{\textit{Scenario B.M.2}}\tabularnewline
\cline{2-5} \cline{3-5} \cline{4-5} \cline{5-5} \cline{7-10} \cline{8-10} \cline{9-10} \cline{10-10} 
 & \multicolumn{4}{c}{$n=500,p=84$} &  & \multicolumn{4}{c}{$n=1000,p=84$}\tabularnewline
\cline{2-10} \cline{3-10} \cline{4-10} \cline{5-10} \cline{6-10} \cline{7-10} \cline{8-10} \cline{9-10} \cline{10-10} 
 & $r_{\text{miss}}$ & rec & prec & $F_{1}$ &  & $r_{\text{miss}}$ & rec & prec & $F_{1}$\tabularnewline
\hline 
DART-20 & 0.95 & 0.75(0.01) & 0.91(0.01) & 0.82(0.01) &  & 0.62 & 0.89(0.01) & 0.93(0.01) & 0.91(0.01)\tabularnewline
DART-200 & 0.98 & 0.75(0.01) & 0.32(0.01) & 0.44(0.01) &  & 0.56 & 0.9(0.01) & 0.52(0.02) & 0.65(0.01)\tabularnewline
ABC-10-0.25 & 1 & 0.71(0.01) & 0.85(0.01) & 0.76(0.01) &  & 1 & 0.81(0.01) & 0.93(0.01) & 0.86(0.01)\tabularnewline
ABC-10-0.50 & 1 & 0.53(0.01) & 1(0) & 0.69(0.01) &  & 1 & 0.71(0.01) & 1(0) & 0.83(0.01)\tabularnewline
ABC-20-0.25 & 0.98 & 0.8(0.01) & 0.22(0) & 0.34(0) &  & 0.78 & 0.87(0.01) & 0.33(0.01) & 0.48(0.01)\tabularnewline
ABC-20-0.50 & 1 & 0.62(0.01) & 0.96(0.01) & 0.75(0.01) &  & 1 & 0.77(0.01) & 0.98(0.01) & 0.86(0.01)\tabularnewline
Permute (VIP) & 0.93 & 0.8(0.01) & 0.81(0.01) & 0.8(0.01) &  & 0.28 & 0.95(0.01) & 0.91(0.01) & 0.93(0.01)\tabularnewline
Permute (Within-Type VIP) & 0.96 & 0.79(0.01) & 0.83(0.02) & 0.8(0.01) &  & 0.52 & 0.91(0.01) & 0.87(0.02) & 0.88(0.01)\tabularnewline
Permute (MI) & 0.94 & 0.8(0.01) & 0.79(0.01) & 0.79(0.01) &  & 0.36 & 0.94(0.01) & 0.84(0.01) & 0.88(0.01)\tabularnewline
Backward & 0.67 & 0.84(0.01) & 0.54(0.03) & 0.57(0.03) &  & 0.15 & 0.97(0.01) & 0.55(0.03) & 0.63(0.03)\tabularnewline
\hline 
\end{tabular}
}
\end{table}

Simulation results are shown in Table \ref{tab:bYmX}. In presence
of binary responses and mixed-type predictors, the variable selection
results are clearly improved by increasing the sample size from $n=500$
to $n=1000$. Under the independent setting (Scenario B.M.1),
the three proposed approaches along with the approach of \citet{bleich2014variable}
achieve $r_{\text{miss}}=0$ and $\text{rec}\equiv1$ for both
$p=50$ and $p=200$, with $n=1000$ samples. DART with $M=200$ trees
gets slightly worse $r_{\text{miss}}$ and recall scores under these
two settings. However, when correlation is introduced, none of the
methods work very well in identifying all the relevant predictors,
though the backward selection procedure still gets the best recall
and $r_{\text{miss}}$ scores. In fact, we find that the result of
the backward selection procedure can be greatly improved by thinning
the MCMC chain of BART. For the Scenario B.M.2 with $n=1000$,
we get $r_{\text{miss}}=0.06$, $\text{rec}=0.99(0.01)$, $\text{prec}=0.67(0.05)$
and $F_{1}=0.73(0.05)$ by keeping every $10^{\text{th}}$ posterior
sample in each BART run.

\subsection{Conclusion of Simulation Results}

We summarize the simulation results above and in Section S.4 of 
the Supplementary Material in Table \ref{tab:summary}. We
call a variable selection approach with $r_{\text{miss}}$ no greater
than $0.1$ and precision no less than $0.6$, i.e., an approach with
excellent capability of including all the relevant predictors and
acceptable capability of excluding irrelevant predictors, a successful
approach, and check it in Table \ref{tab:summary}. From the table,
we can see that the backward selection approach achieves the highest
success rate, followed by the two new permutation-based approaches,
meaning that the three proposed approaches consistently perform well
in identifying all the relevant predictors and excluding irrelevant
predictors. A drawback of the three proposed approaches is that, like
existing BART-based variable selection approaches, they also suffer
from multicollinearity (Scenario C.M.2, Scenario B.C.2 and
Scenario B.M.2), especially when the noise is high or the response
is binary. Another shortcoming of the backward selection approach
is the computational cost of running BART multiple times, but it should
be noted that at each step of the backward selection, BART models
can be fitted in parallel on multiple cores.

\begin{table}[ht]
\caption{Summary of the simulation studies based on Table \ref{tab:cYcX}--\ref{tab:bYmX} and Table S.1--S.2 of the Supplementary Material. Variable selection
approaches with $r_{\text{miss}}$ no greater than $0.1$ and precision
no less than $0.6$ are checked.}
\label{tab:summary}

\resizebox*{\textwidth}{!}{
\begin{tabular}{lllcccccccccc}
\hline 
\multicolumn{2}{c}{Evaluated Setting} &  & \multicolumn{10}{c}{Evaluated Variable Selection Approaches}\tabularnewline
\cline{1-2} \cline{2-2} \cline{4-13} \cline{5-13} \cline{6-13} \cline{7-13} \cline{8-13} \cline{9-13} \cline{10-13} \cline{11-13} \cline{12-13} \cline{13-13} 
Scenario & $(n,p,\sigma^{2})$ &  & DART-20 & DART-200 & ABC-10-0.25 & ABC-10-0.50 & ABC-20-0.25 & ABC-20-0.50 & Permute (VIP) & Permute (Within-Type VIP) & Permute (MI) & Backward\tabularnewline
\hline
C.C.1 & $(500,50,1)$ &  & $\checkmark$ & $\checkmark$ & $\checkmark$ & $\checkmark$ &  & $\checkmark$ & $\checkmark$ & $-$ & $\checkmark$ & $\checkmark$\tabularnewline
\hline 
C.C.1 & $(500,200,1)$ &  & $\checkmark$ & $\checkmark$ & $\checkmark$ &  & $\checkmark$ & $\checkmark$ & $\checkmark$ & $-$ & $\checkmark$ & $\checkmark$\tabularnewline
\hline 
C.C.2 & $(500,50,1)$ &  &  & $\checkmark$ & $\checkmark$ &  &  & $\checkmark$ & $\checkmark$ & $-$ & $\checkmark$ & $\checkmark$\tabularnewline
\hline 
C.C.2 & $(500,200,1)$ &  &  &  &  &  &  &  &  & $-$ &  & $\checkmark$\tabularnewline
\hline 
C.M.1 & $(500,50,1)$ &  & $\checkmark$ & $\checkmark$ & $\checkmark$ & $\checkmark$ & $\checkmark$ & $\checkmark$ &  & $\checkmark$ & $\checkmark$ & $\checkmark$\tabularnewline
\hline 
C.M.1 & $(500,200,1)$ &  & $\checkmark$ & $\checkmark$ & $\checkmark$ &  & $\checkmark$ & $\checkmark$ & $\checkmark$ & $\checkmark$ & $\checkmark$ & $\checkmark$\tabularnewline
\hline 
C.M.1 & $(1000,50,1)$ &  & $\checkmark$ & $\checkmark$ & $\checkmark$ & $\checkmark$ & $\checkmark$ & $\checkmark$ &  & $\checkmark$ & $\checkmark$ & $\checkmark$\tabularnewline
\hline 
C.M.1 & $(1000,200,1)$ &  & $\checkmark$ & $\checkmark$ & $\checkmark$ & $\checkmark$ & $\checkmark$ & $\checkmark$ & $\checkmark$ & $\checkmark$ & $\checkmark$ & $\checkmark$\tabularnewline
\hline 
C.M.1 & $(1000,50,10)$ &  & $\checkmark$ & $\checkmark$ & $\checkmark$ & $\checkmark$ &  & $\checkmark$ &  & $\checkmark$ & $\checkmark$ & $\checkmark$\tabularnewline
\hline 
C.M.1 & $(1000,200,10)$ &  & $\checkmark$ &  & $\checkmark$ &  & $\checkmark$ & $\checkmark$ & $\checkmark$ & $\checkmark$ & $\checkmark$ & $\checkmark$\tabularnewline
\hline 
C.M.2 & $(500,84,1)$ &  &  &  &  &  &  &  &  &  & $\checkmark$ & \tabularnewline
\hline 
C.M.2 & $(500,84,10)$ &  &  &  &  &  &  &  &  &  &  & \tabularnewline
\hline 
C.M.2 & $(1000,84,1)$ &  & $\checkmark$ &  &  &  & $\checkmark$ &  &  & $\checkmark$ & $\checkmark$ & $\checkmark$\tabularnewline
\hline 
C.M.2 & $(1000,84,10)$ &  &  &  &  &  &  &  &  &  &  & \tabularnewline
\hline 
B.C.1 & $(500,50,-)$ &  & $\checkmark$ & $\checkmark$ & $\checkmark$ &  &  & $\checkmark$ & $\checkmark$ & $-$ & $\checkmark$ & $\checkmark$\tabularnewline
\hline 
B.C.1 & $(500,200,-)$ &  &  &  &  &  &  &  & $\checkmark$ & $-$ & $\checkmark$ & $\checkmark$\tabularnewline
\hline 
B.C.2 & $(500,50,-)$ &  &  &  &  &  &  &  &  & $-$ &  & $\checkmark$\tabularnewline
\hline 
B.C.2 & $(500,200,-)$ &  &  &  &  &  &  &  &  & $-$ &  & \tabularnewline
\hline 
B.M.1 & $(500,50,-)$ &  &  &  &  &  &  &  & $\checkmark$ & $\checkmark$ & $\checkmark$ & $\checkmark$\tabularnewline
\hline 
B.M.1 & $(500,200,-)$ &  &  &  &  &  &  &  &  &  &  & \tabularnewline
\hline 
B.M.1 & $(1000,50,-)$ &  & $\checkmark$ & $\checkmark$ & $\checkmark$ &  &  & $\checkmark$ & $\checkmark$ & $\checkmark$ & $\checkmark$ & $\checkmark$\tabularnewline
\hline 
B.M.1 & $(1000,200,-)$ &  &  & $\checkmark$ &  &  &  &  & $\checkmark$ & $\checkmark$ & $\checkmark$ & $\checkmark$\tabularnewline
\hline 
B.M.2 & $(500,84,-)$ &  &  &  &  &  &  &  &  &  &  & \tabularnewline
\hline 
B.M.2 & $(1000,84,-)$ &  &  &  &  &  &  &  &  &  &  & \tabularnewline
\hline 
\multicolumn{2}{l}{Success rate} &  & $45.8\%$ & $45.8\%$ & $45.8\%$ & $20.8\%$ & $29.2\%$ & $45.8\%$ & $45.8\%$ & \textbf{$\boldsymbol{62.5\%}$} & \textbf{$\boldsymbol{66.7\%}$} & \textbf{$\boldsymbol{70.8\%}$}\tabularnewline
\hline 
\end{tabular}
}
\end{table}

\section{Discussion and Future Work\label{sec:5}}

This paper reviews and explores existing BART-based variable selection
methods and introduces three new variable selection approaches. These
new approaches are designed for adapting to data with mixed-type predictors,
and more importantly, for better allowing all the relevant predictors
into the model.

We outline some interesting areas for future work. First, the distribution
of the null importance scores in the permutation-based approach is
unknown. If the distribution can be approximated appropriately, then
a better threshold can be used. Second, from the simulation studies
for the backward selection approach, we note that there is only a
slight difference in $\text{MSE}_{\text{test}}$ between the winner
models from two consecutive steps, when both of them include all the
relevant predictors. However, if one of the winner models drops a
relevant predictor, the difference can become very large. In light
of this, it would be advantageous to develop a formal stopping rule,
which would also improve the precision score and the efficiency of
the backward selection approach. Finally, a potential direction of
conducting variable selection based on BART is to take further advantage
of the model selection property of BART itself. As discussed in Section
\ref{subsec:2_1}, BART can be regarded as a Bayesian model selection approach,  but it does not perform well due to a large number of noise predictors. We may
alleviate this issue by muting noise predictors adaptively and externally
(\citet{zhu2015reinforcement}).

\section*{Acknowledgements}
Luo and Daniels were partially supported by NIH R01CA 183854.

\printbibliography

\clearpage

\begin{center}
\LARGE Supplement to "Variable Selection Using Bayesian Additive Regression Trees"
\end{center}

\beginsupplement

\section{Proof of Lemma 3.2}

\begin{proof}
From Cauchy-Schwarz inequality, Definition 2.1 and 
Definition 3.1 of the main paper, it is clear that
\begin{eqnarray*}
\left|\tilde{v}_{j}-v_{j}\right| & = & \left|\sum\limits _{k=1}^{K}\left(c_{jk}\frac{c_{\cdot k}-\bar{c}_{\cdot K}}{K\bar{c}_{\cdot K}c_{\cdot k}}\right)\right|  \cdot \sum\limits _{k=1}^{K}\left(\frac{1}{K}\frac{c_{jk}}{c_{\cdot k}}\frac{\left|c_{\cdot k}-\bar{c}_{\cdot K}\right|}{\bar{c}_{\cdot K}}\right)\\
& \leq & \left[\frac{1}{K}\sum\limits _{k=1}^{K}\left(\frac{c_{jk}}{c_{\cdot k}}\right)^{2}\right]^{1/2} \cdot \frac{\left[\sum\limits _{k=1}^{K}\left(c_{\cdot k}-\bar{c}_{\cdot K}\right)^{2} / K \right]^{1/2}}{\bar{c}_{\cdot K}} \\
& \le & \delta_{1}^{1/2}\delta_{2}.
\end{eqnarray*}
\end{proof}

\section{Decomposition of Untruncated Metropolis Ratio}

The untruncated Metropolis ratio in Equation (9) of the main paper 
can be explicitly expressed as
\begin{align*}
r(\eta) = & \frac{P\left(T\mid T^{*}\right)}{P\left(T^{*}\mid T\right)} \cdot \frac{P\left(T^{*}\right)}{P\left(T\right)} \cdot \frac{P\left(\boldsymbol{r}\mid T^{*},\sigma^{2}\right)}{P\left(\boldsymbol{r}\mid T,\sigma^{2}\right)} \\
= & \frac{P\left(\text{DEATH}\right) / w_{2}^{*}}{P\left(\text{BIRTH}\right) / \left[b \cdot p_{\text{adj}}(\eta) \cdot n_{j,\text{adj}}(\eta) \right]} \cdot   \frac{\gamma \cdot \left[1 - \gamma / (2 + d_{\eta})^{\beta} \right]^{2}}{\left[(1+d_{\eta})^{\beta} - \gamma \right] \cdot p_{\text{adj}}(\eta) \cdot n_{j,\text{adj}}(\eta) }  \cdot  \frac{P\left(\boldsymbol{r}\mid T^{*},\sigma^{2}\right)}{P\left(\boldsymbol{r}\mid T,\sigma^{2}\right)}\\
= & \frac{b}{w_{2}^{*}} \cdot \frac{\gamma \cdot \left[1 - \gamma / (2 + d_{\eta})^{\beta} \right]^{2}}{(1+d_{\eta})^{\beta}-\gamma} \cdot \frac{P\left(\boldsymbol{r}\mid T^{*},\sigma^{2}\right)}{P\left(\boldsymbol{r}\mid T,\sigma^{2}\right)} \\
= & \frac{2b}{b+2} \cdot \frac{\gamma \cdot \left[1 - \gamma / (2 + d_{\eta})^{\beta} \right]^{2}}{(1+d_{\eta})^{\beta}-\gamma} \cdot \frac{P\left(\boldsymbol{r}\mid T^{*},\sigma^{2}\right)}{P\left(\boldsymbol{r}\mid T,\sigma^{2}\right)}, 
\end{align*}
where $w_{2}^{*}$ is the number of second generation internal nodes
(i.e., nodes with two terminal child nodes) in the proposed tree $T^{*}$,
$b$ is the number of terminal nodes in the current tree $T$, $p_{\text{adj}}(\eta)$
is the number of predictors available to split on at the node $\eta$
in the current tree $T$, $n_{j,\text{adj}}(\eta)$ is the number
of available values to choose for the selected predictor $x_{j}$
at the node $\eta$ in the current tree $T$, and $d_{\eta}$ is the
depth of node $\eta$ in the current tree $T$. 

The second equality is because the proposed tree $T^{*}$ is identical
to the current tree $T$ except for the new split node. The third
equality is from that $P(\text{BIRTH})=P(\text{DEATH})=0.5$. The
last equality is due to the properties of a binary tree, $b=2w_{2}$
and $w_{2}^{*}=w_{2}+1$, where $w_{2}$ is the number of second generation
internal nodes in the current tree $T$.

\section{Brief Introduction to BartMixVs}

Built upon the CRAN \proglang{R} package \pkg{BART} (\citet{sparapani2021nonparametric}), the \proglang{R} package \pkg{BartMixVs} (\citet{luo2021package}) is developed to implement the three proposed variable selection approaches of the main paper and three existing BART-based variable selection approaches: the permutation-based variable selection approach using BART VIP (\citet{bleich2014variable}), DART (\citet{linero2018bayesian}) and ABC Bayesian forest (\citet{liu2019variable}). The simulation results of this work were obtained using this package which is available at \url{https://github.com/chujiluo/BartMixVs}.

\section{More Simulations}

\subsection{Additional Simulations for Continuous Response and Mixed-Type Predictors}

In this section, we provide additional simulation results for Scenario C.M.1 and Scenario C.M.2 of the main paper, as shown in Table \ref{tab:cYmX-2}.

\subsection{Simulations for Binary Response and Continuous Predictors}

In this setting, we consider the following two scenarios with $n=500$
and $p\in\{50,200\}$.

\begin{scenarioBC1}
Sample the predictors $x_{1},\cdots,x_{p}$ from
$\text{Uniform}(0,1)$ independently and the response $y$ from $\text{Bernoulli}(\Phi(f_{0}(\boldsymbol{x}))$,
where $f_{0}(x)$ is defined in (12) of the main paper.
\end{scenarioBC1}

\begin{scenarioBC2}
Sample the predictors $x_{1},\cdots,x_{p}$ from
$\text{Normal}(0,\Sigma)$ with $\Sigma_{jk}=0.3^{|j-k|}$, $j,k=1,\cdots,p$,
and the response $y$ from $\text{Bernoulli}(\Phi(f_{0}(\boldsymbol{x}))$, where $f_{0}(\boldsymbol{x})$ is defined in (13) of the main paper.
\end{scenarioBC2}

Table \ref{tab:bYcX} shows the simulation results. When the predictors
are independent (Scenario B.C.1) and $p=50$, all the methods
except ABC-10-0.50 show similar performance. As the dimension increases,
the two permutation-based approaches and the backward selection approach
significantly outperform other methods in terms of the $r_{\text{miss}}$
and recall scores. When correlation is introduced (Scenario
B.C.2), the backward selection approach is the best in all metrics.
Moreover, it is the only method that identifies all the relevant predictors
over all the replications when $p=50$.

\begin{table}[ht]
\caption{Additional simulation results for Scenario C.M.1 where predictors are of mixed type and independent, and Scenario C.M.2 where predictors are of mixed type and correlated.}
\label{tab:cYmX-2}
\resizebox*{\textwidth}{!}{
\begin{tabular}{lccccccccc}
\hline 
 & \multicolumn{9}{c}{Scenario C.M.1}\tabularnewline
\cline{2-10} \cline{3-10} \cline{4-10} \cline{5-10} \cline{6-10} \cline{7-10} \cline{8-10} \cline{9-10} \cline{10-10} 
 & \multicolumn{4}{c}{\textit{$n=1000,p=50,\sigma^{2}=1$}} &  & \multicolumn{4}{c}{\textit{$n=1000,p=200,\sigma^{2}=1$}}\tabularnewline
\cline{2-5} \cline{3-5} \cline{4-5} \cline{5-5} \cline{7-10} \cline{8-10} \cline{9-10} \cline{10-10} 
 & $r_{\text{miss}}$ & rec & prec & $F_{1}$ &  & $r_{\text{miss}}$ & rec & prec & $F_{1}$\tabularnewline
\hline 
DART-20 & 0 & 1(0) & 0.93(0.01) & 0.96(0.01) &  & 0 & 1(0) & 0.91(0.01) & 0.95(0.01)\tabularnewline
DART-200 & 0 & 1(0) & 0.6(0.01) & 0.75(0.01) &  & 0 & 1(0) & 0.65(0.02) & 0.78(0.01)\tabularnewline
ABC-10-0.25 & 0 & 1(0) & 1(0) & 1(0) &  & 0 & 1(0) & 1(0) & 1(0)\tabularnewline
ABC-10-0.50 & 0 & 1(0) & 1(0) & 1(0) &  & 0 & 1(0) & 1(0) & 1(0)\tabularnewline
ABC-20-0.25 & 0 & 1(0) & 0.79(0.01) & 0.88(0.01) &  & 0 & 1(0) & 0.98(0.01) & 0.99(0)\tabularnewline
ABC-20-0.50 & 0 & 1(0) & 1(0) & 1(0) &  & 0 & 1(0) & 1(0) & 1(0)\tabularnewline
Permute (VIP) & 1 & 0.76(0.01) & 1(0) & 0.86(0.01) &  & 0 & 1(0) & 0.94(0.01) & 0.97(0.01)\tabularnewline
Permute (Within-Type VIP) & 0 & 1(0) & 0.99(0.01) & 0.99(0) &  & 0 & 1(0) & 0.69(0.02) & 0.81(0.01)\tabularnewline
Permute (MI) & 0 & 1(0) & 0.98(0.01) & 0.99(0) &  & 0 & 1(0) & 0.89(0.01) & 0.94(0.01)\tabularnewline
Backward & 0 & 1(0) & 0.8(0.02) & 0.87(0.01) &  & 0 & 1(0) & 0.8(0.02) & 0.87(0.01)\tabularnewline
\cline{2-10} \cline{3-10} \cline{4-10} \cline{5-10} \cline{6-10} \cline{7-10} \cline{8-10} \cline{9-10} \cline{10-10} 
 & \multicolumn{4}{c}{\textit{$n=1000,p=50,\sigma^{2}=10$}} &  & \multicolumn{4}{c}{\textit{$n=1000,p=200,\sigma^{2}=10$}}\tabularnewline
\cline{2-5} \cline{3-5} \cline{4-5} \cline{5-5} \cline{7-10} \cline{8-10} \cline{9-10} \cline{10-10} 
 & $r_{\text{miss}}$ & rec & prec & $F_{1}$ &  & $r_{\text{miss}}$ & rec & prec & $F_{1}$\tabularnewline
\hline 
DART-20 & 0 & 1(0) & 0.98(0.01) & 0.99(0) &  & 0 & 1(0) & 0.96(0.01) & 0.97(0.01)\tabularnewline
DART-200 & 0 & 1(0) & 0.67(0.02) & 0.79(0.01) &  & 0 & 1(0) & 0.54(0.02) & 0.69(0.01)\tabularnewline
ABC-10-0.25 & 0 & 1(0) & 0.97(0.01) & 0.98(0) &  & 0 & 1(0) & 0.99(0) & 1(0)\tabularnewline
ABC-10-0.50 & 0 & 1(0) & 1(0) & 1(0) &  & 0.19 & 0.96(0.01) & 1(0) & 0.98(0)\tabularnewline
ABC-20-0.25 & 0 & 1(0) & 0.45(0.01) & 0.61(0.01) &  & 0 & 1(0) & 0.92(0.01) & 0.96(0.01)\tabularnewline
ABC-20-0.50 & 0 & 1(0) & 0.99(0) & 0.99(0) &  & 0 & 1(0) & 1(0) & 1(0)\tabularnewline
Permute (VIP) & 0.11 & 0.98(0.01) & 0.98(0.01) & 0.98(0) &  & 0 & 1(0) & 0.69(0.01) & 0.81(0.01)\tabularnewline
Permute (Within-Type VIP) & 0 & 1(0) & 0.95(0.01) & 0.97(0.01) &  & 0 & 1(0) & 0.64(0.02) & 0.77(0.01)\tabularnewline
Permute (MI) & 0 & 1(0) & 0.84(0.01) & 0.91(0.01) &  & 0 & 1(0) & 0.62(0.01) & 0.76(0.01)\tabularnewline
Backward & 0 & 1(0) & 0.68(0.03) & 0.78(0.02) &  & 0 & 1(0) & 0.60(0.02) & 0.74(0.02)\tabularnewline
\hline 
 & \multicolumn{9}{c}{Scenario C.M.2}\tabularnewline
\cline{2-10} \cline{3-10} \cline{4-10} \cline{5-10} \cline{6-10} \cline{7-10} \cline{8-10} \cline{9-10} \cline{10-10} 
 & \multicolumn{4}{c}{$n=500,p=84,\sigma^{2}=10$} &  & \multicolumn{4}{c}{$n=1000,p=84,\sigma^{2}=10$}\tabularnewline
\cline{2-5} \cline{3-5} \cline{4-5} \cline{5-5} \cline{7-10} \cline{8-10} \cline{9-10} \cline{10-10} 
 & $r_{\text{miss}}$ & rec & prec & $F_{1}$ &  & $r_{\text{miss}}$ & rec & prec & $F_{1}$\tabularnewline
\hline 
DART-20 & 1 & 0.5(0.02) & 0.75(0.02) & 0.58(0.01) &  & 0.95 & 0.68(0.01) & 0.85(0.02) & 0.75(0.01)\tabularnewline
DART-200 & 1 & 0.61(0.01) & 0.26(0.01) & 0.36(0.01) &  & 0.91 & 0.77(0.01) & 0.36(0.01) & 0.48(0.01)\tabularnewline
ABC-10-0.25 & 1 & 0.52(0.01) & 0.7(0.02) & 0.58(0.01) &  & 1 & 0.64(0.01) & 0.84(0.02) & 0.72(0.01)\tabularnewline
ABC-10-0.50 & 1 & 0.28(0.01) & 0.98(0.01) & 0.43(0.01) &  & 1 & 0.43(0.01) & 0.99(0) & 0.59(0.01)\tabularnewline
ABC-20-0.25 & 0.99 & 0.7(0.01) & 0.18(0) & 0.29(0.01) &  & 0.93 & 0.79(0.01) & 0.31(0.01) & 0.45(0.01)\tabularnewline
ABC-20-0.50 & 1 & 0.39(0.01) & 0.89(0.02) & 0.53(0.01) &  & 1 & 0.55(0.01) & 0.97(0.01) & 0.7(0.01)\tabularnewline
Permute (VIP) & 0.99 & 0.62(0.02) & 0.67(0.02) & 0.64(0.01) &  & 0.91 & 0.79(0.01) & 0.77(0.02) & 0.77(0.01)\tabularnewline
Permute (Within-Type VIP) & 1 & 0.61(0.02) & 0.69(0.03) & 0.64(0.02) &  & 0.98 & 0.75(0.02) & 0.79(0.02) & 0.76(0.01)\tabularnewline
Permute (MI) & 0.99 & 0.62(0.01) & 0.65(0.02) & 0.63(0.01) &  & 0.91 & 0.8(0.01) & 0.73(0.02) & 0.75(0.01)\tabularnewline
Backward & 0.75 & 0.65(0.03) & 0.36(0.03) & 0.36(0.02) &  & 0.67 & 0.81(0.02) & 0.31(0.03) & 0.37(0.02)\tabularnewline
\hline 
\end{tabular}
}
\end{table}

\begin{table}[!htb]
\caption{Simulation results for the scenarios using binary
responses and continuous predictors. Predictors in Scenario
B.C.1 are independent and predictors in Scenario B.C.2 are
correlated.}
\label{tab:bYcX}
\resizebox*{\textwidth}{!}{
\begin{tabular}{lccccccccc}
\hline 
 & \multicolumn{9}{c}{Scenario B.C.1}\tabularnewline
\cline{2-10} \cline{3-10} \cline{4-10} \cline{5-10} \cline{6-10} \cline{7-10} \cline{8-10} \cline{9-10} \cline{10-10} 
 & \multicolumn{4}{c}{\textit{$n=500,p=50$}} &  & \multicolumn{4}{c}{\textit{$n=500,p=200$}}\tabularnewline
\cline{2-5} \cline{3-5} \cline{4-5} \cline{5-5} \cline{7-10} \cline{8-10} \cline{9-10} \cline{10-10} 
 & $r_{\text{miss}}$ & rec & prec & $F_{1}$ &  & $r_{\text{miss}}$ & rec & prec & $F_{1}$\tabularnewline
\hline 
DART-20 & 0 & 1(0) & 0.98(0.01) & 0.99(0) &  & 0.28 & 0.94(0.01) & 0.98(0.01) & 0.96(0.01)\tabularnewline
DART-200 & 0 & 1(0) & 0.97(0.01) & 0.98(0) &  & 0.11 & 0.98(0.01) & 0.96(0.01) & 0.97(0.01)\tabularnewline
ABC-10-0.25 & 0 & 1(0) & 0.95(0.01) & 0.97(0) &  & 0.67 & 0.86(0.01) & 1(0) & 0.92(0.01)\tabularnewline
ABC-10-0.50 & 0.33 & 0.93(0.01) & 1(0) & 0.96(0.01) &  & 0.99 & 0.74(0.01) & 1(0) & 0.84(0.01)\tabularnewline
ABC-20-0.25 & 0 & 1(0) & 0.17(0) & 0.29(0) &  & 0.15 & 0.97(0.01) & 0.98(0.01) & 0.97(0.01)\tabularnewline
ABC-20-0.50 & 0.03 & 0.99(0) & 1(0) & 0.99(0) &  & 0.91 & 0.8(0.01) & 1(0) & 0.88(0.01)\tabularnewline
Permute (VIP) & 0 & 1(0) & 1(0) & 1(0) &  & 0 & 1(0) & 0.88(0.01) & 0.94(0.01)\tabularnewline
Permute (MI) & 0 & 1(0) & 0.96(0.01) & 0.98(0) &  & 0 & 1(0) & 0.91(0.01) & 0.95(0.01)\tabularnewline
Backward & 0.03 & 0.99(0) & 0.92(0.02) & 0.94(0.02) &  & 0.05 & 0.99(0) & 0.94(0.02) & 0.95(0.01)\tabularnewline
\hline
 & \multicolumn{9}{c}{Scenario B.C.2}\tabularnewline
\cline{2-10} \cline{3-10} \cline{4-10} \cline{5-10} \cline{6-10} \cline{7-10} \cline{8-10} \cline{9-10} \cline{10-10} 
 & \multicolumn{4}{c}{$n=500,p=50$} &  & \multicolumn{4}{c}{$n=500,p=200$}\tabularnewline
\cline{2-5} \cline{3-5} \cline{4-5} \cline{5-5} \cline{7-10} \cline{8-10} \cline{9-10} \cline{10-10} 
 & $r_{\text{miss}}$ & rec & prec & $F_{1}$ &  & $r_{\text{miss}}$ & rec & prec & $F_{1}$\tabularnewline
\hline 
DART-20 & 0.87 & 0.49(0.03) & 0.71(0.04) & 0.57(0.03) &  & 1 & 0.06(0.02) & 0.09(0.02) & 0.07(0.02)\tabularnewline
DART-200 & 0.42 & 0.78(0.03) & 0.67(0.03) & 0.69(0.02) &  & 0.92 & 0.35(0.03) & 0.09(0.01) & 0.11(0.02)\tabularnewline
ABC-10-0.25 & 0.46 & 0.82(0.02) & 0.54(0.02) & 0.64(0.02) &  & 1 & 0(0) & 0.01(0.01) & 0(0)\tabularnewline
ABC-10-0.50 & 1 & 0.05(0.01) & 0.14(0.04) & 0.07(0.02) &  & 1 & 0(0) & 0(0) & 0(0)\tabularnewline
ABC-20-0.25 & 0 & 1(0) & 0.1(0) & 0.18(0) &  & 1 & 0.01(0) & 0.02(0.01) & 0.01(0.01)\tabularnewline
ABC-20-0.50 & 0.95 & 0.43(0.03) & 0.77(0.04) & 0.53(0.03) &  & 1 & 0(0) & 0(0) & 0(0)\tabularnewline
Permute (VIP) & 0.32 & 0.9(0.02) & 0.85(0.02) & 0.86(0.01) &  & 0.99 & 0.16(0.02) & 0.07(0.01) & 0.1(0.01)\tabularnewline
Permute (MI) & 0.7 & 0.65(0.03) & 0.74(0.02) & 0.67(0.02) &  & 1 & 0.03(0.01) & 0.04(0.01) & 0.03(0.01)\tabularnewline
Backward & 0 & 1(0) & 0.9(0.02) & 0.94(0.01) &  & 0.29 & 0.86(0.02) & 0.62(0.04) & 0.61(0.04)\tabularnewline
\hline 
\end{tabular}
}
\end{table}

\newpage

\section{More Figures}

\begin{figure}[ht]
\centering\includegraphics[width=0.7\columnwidth]{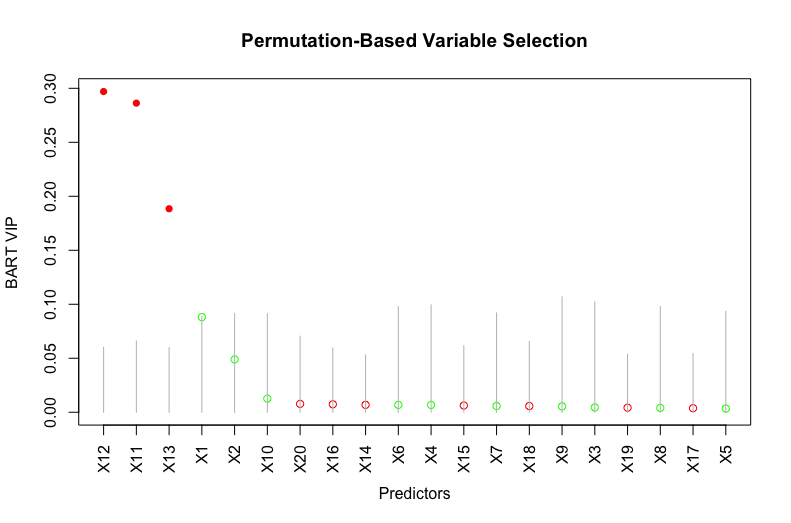}
\caption{\label{fig:Bleich}Variable selection results of \citet{bleich2014variable} 
with the setting, $L=100$, $L_{\text{rep}}=10$,
$\alpha=0.05$ and $M=20$, for Example 2 of the main paper. Red
(or green) dots are for continuous (or binary) predictors. Solid (or
open) dots are for selected (or not selected) predictors. Each vertical
grey line is the local threshold for the corresponding predictor.
Relative binary predictors $x_{1}$ and $x_{2}$ are not identified by this method.}
\end{figure}

\begin{figure}[ht]
\centering%
\begin{minipage}[t]{0.45\textwidth}%
\begin{center}
\includegraphics[width=0.34\paperwidth,height=0.37\paperwidth]{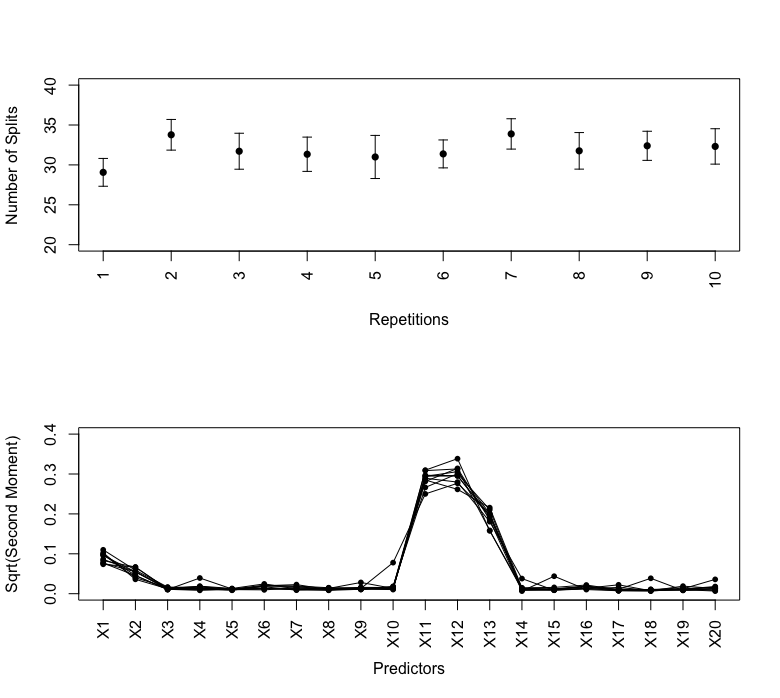}
\par\end{center}%
\end{minipage}%
\begin{minipage}[t]{0.45\textwidth}%
\begin{center}
\includegraphics[width=0.34\paperwidth,height=0.37\paperwidth]{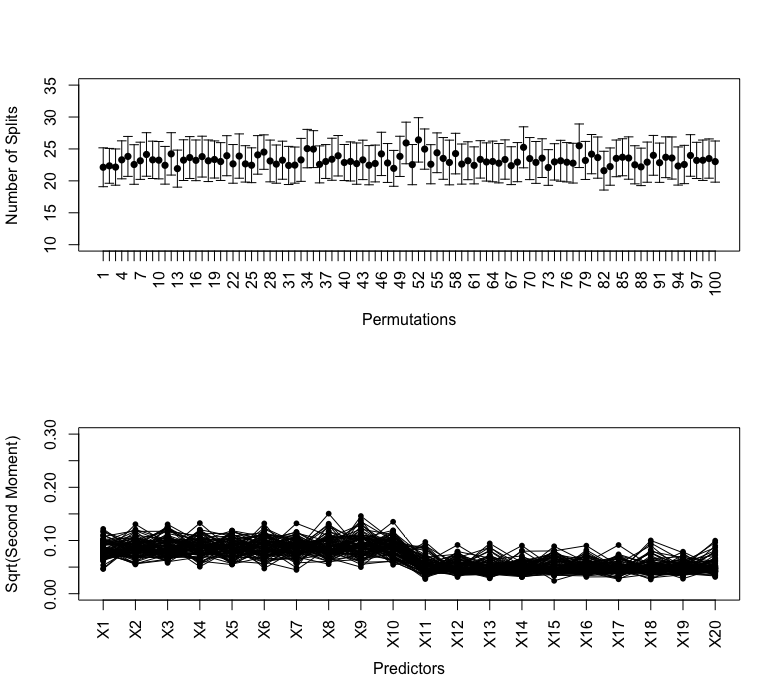}
\par\end{center}%
\end{minipage}
\caption{\label{fig:VIP-tilde}The top two subfigures depict the error
bars of $\{c_{\cdot k}\}_{k=1}^{K}$,
the number of splitting nodes in a posterior sample, for different BART models.
The bottom two subfigures depict $[\sum_{k=1}^{K}(c_{jk}/c_{\cdot k})^{2}/K]^{1/2}$,
the squared root of the second moment of the proportion of the usage
$c_{jk}/c_{\cdot k}$, for different predictors.
Each line is for a BART model.
The posterior samples are obtained from the BART models fitted in 
the approach of \citet{bleich2014variable} applied to Example 2 of the main paper. 
The left two subfigures are based on the $10$ repeated BART models (with different initial values)
on the original dataset. The right two subfigures are based on the $100$
BART models on the null datasets.}
\end{figure}

\begin{figure}[ht]
\centering\includegraphics[width=0.7\columnwidth]{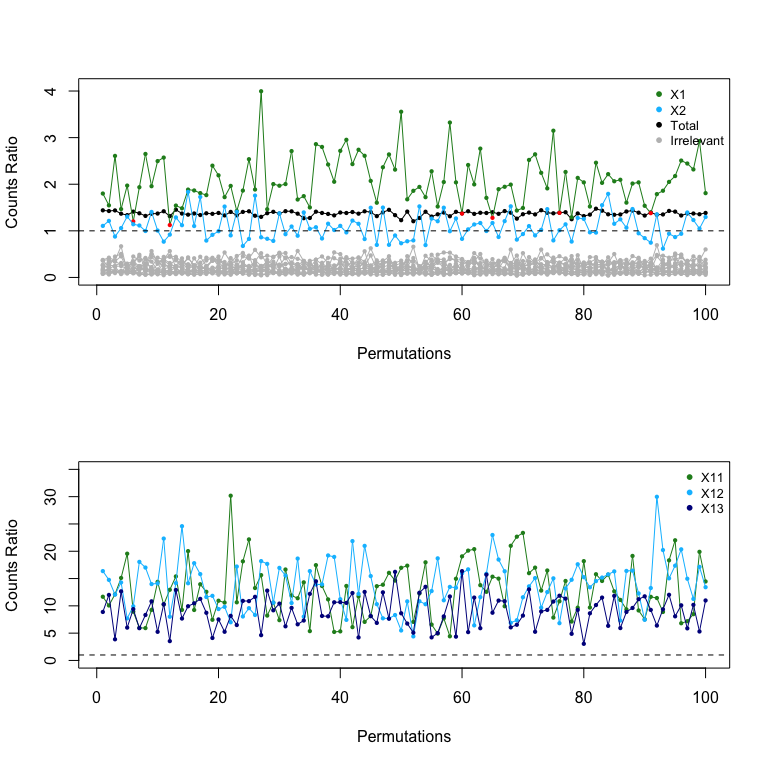}
\caption{\label{fig:Counts-Ratio}The two subfigures depict the overall counts
ratios $\bar{c}_{L_{rep},\cdot\cdot} / c_{l,\cdot\cdot}^{*}$
(black dots) and the counts ratios $\bar{c}_{L_{rep},j\cdot} / c_{l,j\cdot}^{*}$
of a predictor (non-black dots) for $L=100$ permutations. The counts
ratio of the binary predictor $x_{1}$ not satisfying $\bar{c}_{L_{rep},1\cdot} / c_{l,1\cdot}^{*} > \bar{c}_{L_{rep},\cdot\cdot} / c_{l,\cdot\cdot}^{*}$
are marked as red. The posterior samples are obtained from the BART models fitted in 
the approach of \citet{bleich2014variable} applied to Example 2 of the main paper. }
\end{figure}

\begin{figure}[ht]
\centering%
\begin{minipage}[t]{0.45\textwidth}%
\begin{center}
\includegraphics[width=0.3\paperwidth,height=0.24\paperwidth]{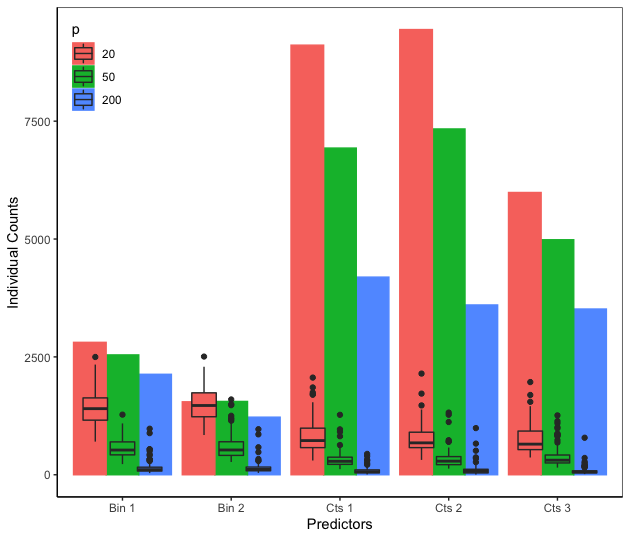}
\par\end{center}%
\end{minipage}%
\begin{minipage}[t]{0.45\textwidth}%
\begin{center}
\includegraphics[width=0.3\paperwidth,height=0.24\paperwidth]{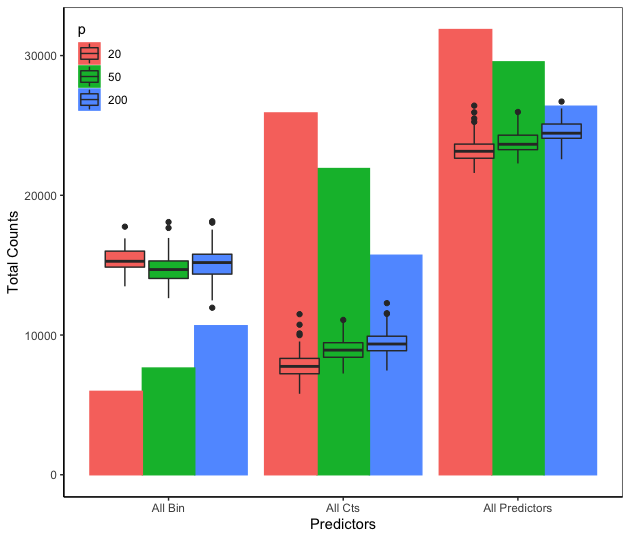}
\par\end{center}%
\end{minipage}
\caption{\label{fig:Counts}The left subfigure depicts $\bar{c}_{L_{\text{rep}},j\cdot}$
(bar) and $\{c_{l,j\cdot}^{*}\}_{l=1}^{L}$ (boxplot) for each relevant
predictor; the right subfigure depicts $\bar{c}_{L_{\text{rep}},\text{cts}\cdot}=(\sum_{r=1}^{L_{\text{rep}}}\sum_{j\in\mathcal{S}_{\text{cts}}}c_{r,j\cdot})/L_{\text{rep}}$,
$\bar{c}_{L_{\text{rep}},\text{bin}\cdot}=(\sum_{r=1}^{L_{\text{rep}}}\sum_{j\in\mathcal{S}_{\text{bin}}}c_{r,j\cdot})/L_{\text{rep}}$,
and $\bar{c}_{L_{\text{rep}},\cdot\cdot}$ (bars), and $\text{\{}c_{l,\text{cts}\cdot}^{*}=\sum_{j\in\mathcal{S}_{\text{cts}}}c_{l,j\cdot}^{*}\}_{l=1}^{L}$,
$\text{\{}c_{l,\text{bin}\cdot}^{*}=\sum_{j\in\mathcal{S}_{\text{bin}}}c_{l,j\cdot}^{*}\}_{l=1}^{L}$,
and $\text{\{}c_{l,\cdot\cdot}^{*}\}_{l=1}^{L}$ (boxplots). The posterior samples are obtained from the BART models fitted in 
the approach of \citet{bleich2014variable} applied to Example \ref{exp:s_1} with
different $p\in \{20,50,100\}$.}
\end{figure}

\begin{figure}[ht]
\centering\includegraphics[width=0.8\columnwidth]{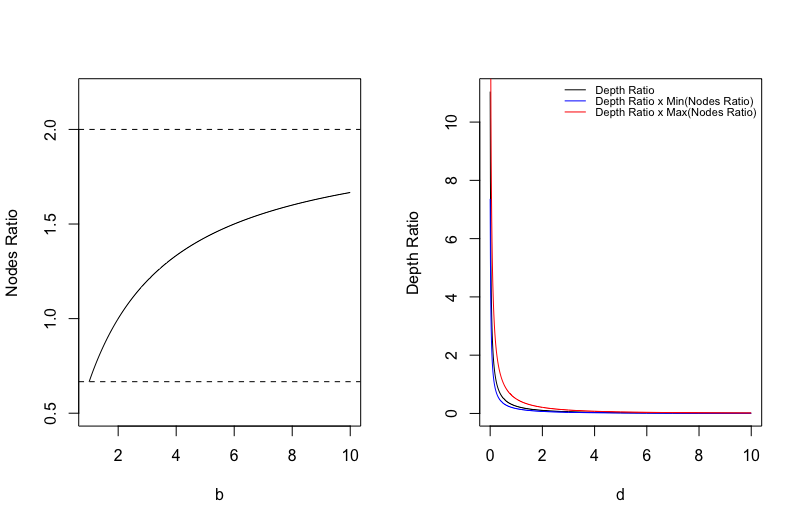}
\caption{\label{fig:Left:-Depth-ratios}The left subfigure depicts nodes ratios
for different $b$'s, the number of terminal nodes; the right
subfigure depicts depth ratios for different depths $d_{\eta}$ when
$\gamma=0.95$ and $\beta=2$.}
\end{figure}

\begin{figure}[ht]
\centering\includegraphics[width=0.7\columnwidth]{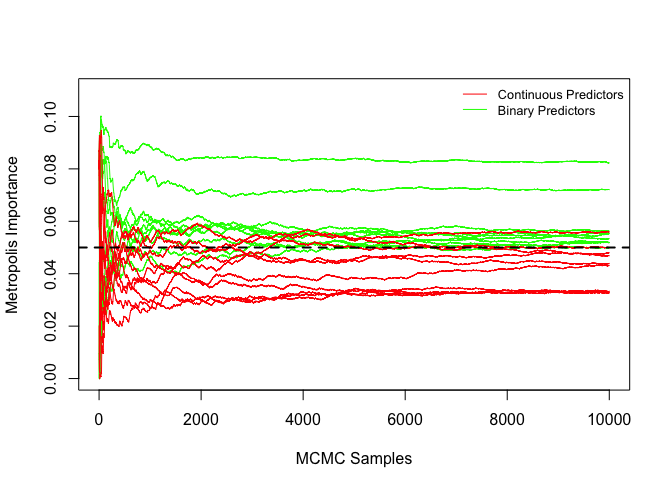}
\caption{\label{fig:MI-Not-Converge}Each line depicts the BART MIs of a predictor
for different numbers of posterior samples (after burning in $1000$
samples) obtained from a BART model built on a null dataset of Example 2 of
the main paper.}
\end{figure}

\begin{figure}[ht]
\centering\includegraphics[width=0.64\columnwidth]{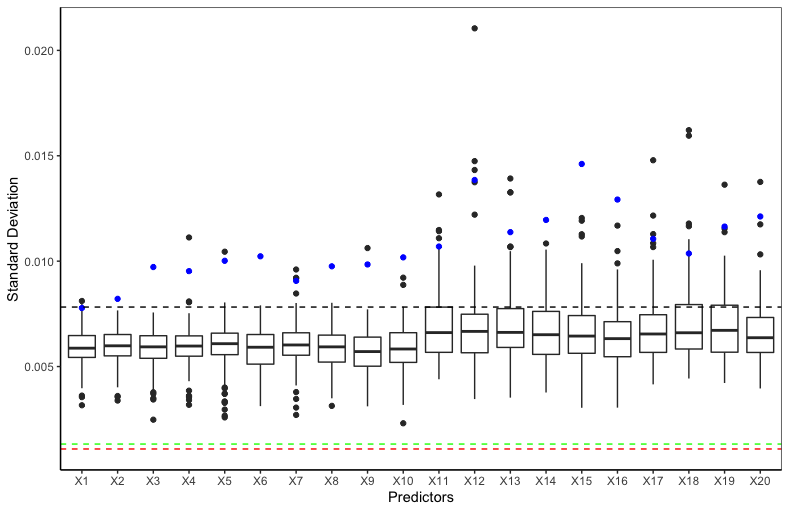}
\caption{\label{fig:sd}Each boxplot depicts $\{s_{ij}\}_{i=1}^{100}$ for
the predictor $x_{j}$. Each blue dot represents $s_{j}$ for the
predictor $x_{j}$. The black dashed line represents $s$. The red
dashed line represent the $s$ within continuous predictors and the
green dashed line represents the $s$ within binary predictors.}
\end{figure}

\begin{figure}[ht]
\centering\includegraphics[width=0.7\columnwidth]{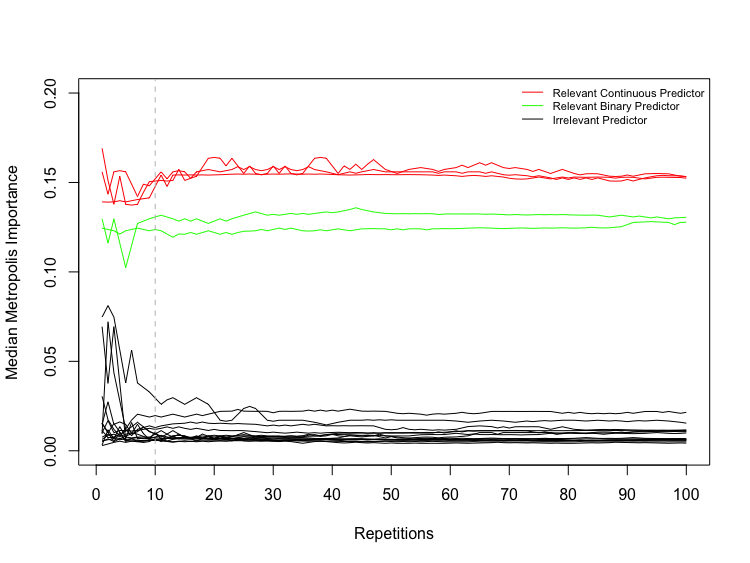}
\caption{\label{fig:Median-of-BART}Each line depicts the median of BART MIs
for a predictor over different numbers of repetitions of BART built on the
data generated from Example 2 of the main paper.}
\end{figure}

In this section, we list Figure \ref{fig:Bleich}, Figure \ref{fig:VIP-tilde},
Figure \ref{fig:Counts-Ratio}, Figure \ref{fig:Counts}, Figure \ref{fig:Left:-Depth-ratios},
Figure \ref{fig:MI-Not-Converge}, Figure \ref{fig:sd} and Figure \ref{fig:Median-of-BART},
which are discussed in the main paper.

Figure \ref{fig:Counts} is based on the following example, a generalized
example of Example 2 of the main paper.

\begin{example}\label{exp:s_1}
For each $i=1, \cdots, 500$, sample $x_{i,1}, \cdots, x_{i,\left\lceil p/2\right\rceil}$ from $\text{Bernoulli(0.5)}$
independently, $x_{i, \left\lceil p/2\right\rceil+1}, \cdots, x_{i,p}$ from $\text{Uniform(0,1)}$ independently and $y_{i}$
from $\mathrm{Normal}(f_{0}(\boldsymbol{x}_{i}),1)$ independently, where
\begin{equation*}
f_{0}(\boldsymbol{x}) = 10\sin(\pi x_{\left\lceil p/2\right\rceil+1}x_{\left\lceil p/2\right\rceil+2}) + 20(x_{\left\lceil p/2\right\rceil+3}-0.5)^{2}+10x_{1}+5x_{2}.
\end{equation*}
\end{example} 

We consider Example \ref{exp:s_1} with $p \in \{20, 50, 100\}$ in Figure \ref{fig:Counts}, where $p=20$
corresponds to Example 2 of the main paper.

\end{document}